\documentclass[usenatbib]{mn2e}

\pdfoutput=1

\usepackage{fixltx2e}
\usepackage{mathptmx}
\usepackage{color}
\usepackage[pdftex]{graphicx}

\newcommand{\Mpc}{\rm\thinspace Mpc}
\newcommand{\kpc}{\rm\thinspace kpc}

\newcommand{\km}{\rm\thinspace km}

\newcommand{\cm}{\rm\thinspace cm}

\newcommand{\cmsq}{\hbox{$\cm^2\,$}}
\newcommand{\cmcu}{\hbox{$\cm^3\,$}}
\newcommand{\pcmcu}{\hbox{$\cm^{-3}\,$}}

\newcommand{\yr}{\rm\thinspace yr}

\newcommand{\s}{\rm\thinspace s}

\newcommand{\Msun}{\hbox{$\rm\thinspace M_{\odot}$}}

\newcommand{\Msunpyr}{\hbox{$\Msun\yr^{-1}\,$}}
\newcommand{\keV}{\rm\thinspace keV}

\newcommand{\erg}{\rm\thinspace erg}

\newcommand{\ergpcmcu}{\hbox{$\erg\cm^{-3}\,$}}

\newcommand{\ergps}{\hbox{$\erg\s^{-1}\,$}}

\newcommand{\ergcmsqps}{\hbox{$\erg\cm^{-2}\s^{-1}\,$}}
\newcommand{\kmps}{\hbox{$\km\s^{-1}\,$}}

\newcommand{\kmpspMpc}{\hbox{$\kmps\Mpc^{-1}$}}
\newcommand{\Zsun}{\hbox{$\thinspace \mathrm{Z}_{\odot}$}}

\newcommand{\psqcm}{\hbox{$\cm^{-2}\,$}}

\begin{document}

\title{Deep high-resolution X-ray spectra from cool-core clusters}

\author[J.S. Sanders et al]
{J.~S. Sanders$^1$\thanks{E-mail: jss@ast.cam.ac.uk},
  A.~C. Fabian$^1$, K.~A. Frank$^2$, J.~R. Peterson$^2$ and H.~R. Russell$^1$\\
  $^1$ Institute of Astronomy, Madingley Road, Cambridge. CB3 0HA\\
  $^2$ Department of Physics, Purdue University, 525 Northwestern Avenue, West Lafayette, IN 47907-2036, USA
}
\maketitle

\begin{abstract}
  We examine deep \emph{XMM-Newton} Reflection Grating Spectrometer (RGS) spectra from the cores of three X-ray bright cool core galaxy clusters, Abell~262, Abell~3581 and HCG~62. Each of the RGS spectra show Fe~\textsc{xvii} emission lines indicating the presence of gas around 0.5~keV. There is no evidence for O~\textsc{vii} emission which would imply gas at still cooler temperatures. The range in detected gas temperature in these objects is a factor of 3.7, 5.6 and 2 for Abell~262, Abell~3581 and HCG~62, respectively. The coolest detected gas only has a volume filling fraction of 6 and 3 per~cent for Abell 262 and Abell 3581, but is likely to be volume filling in HCG~62. \emph{Chandra} spatially resolved spectroscopy confirms the low volume filling fractions of the cool gas in Abell~262 and Abell~3581, indicating this cool gas exists as cold blobs. Any volume heating mechanism aiming to prevent cooling would overheat the surroundings of the cool gas by a factor of 4. If the gas is radiatively cooling below 0.5~keV, it is cooling at a rate at least an order of magnitude below that at higher temperatures in Abell~262 and Abell~3581 and two-orders of magnitude lower in HCG~62. The gas may be cooling non-radiatively through mixing in these cool blobs, where the energy released by cooling is emitted in the infrared. We find very good agreement between smooth particle inference modelling of the cluster and conventional spectral fitting. Comparing the temperature distribution from this analysis with that expected in a cooling flow, there appears to be a even larger break below 0.5 keV as compared with previous empirical descriptions of the deviations of cooling flow models.
\end{abstract}


\section{Introduction}
The hot (few $10^7$K) intracluster medium (ICM) in galaxy clusters is primarily seen in emission in the X-ray waveband. The ICM contains most of the baryonic cluster mass. In a large fraction of clusters of galaxies, observations show that the X-ray surface brightness steeply rises towards their centres (e.g. \citealt{Stewart84}). In addition, spectrally-derived temperature profiles of clusters typically show drops in temperatures by a factor of 2 or 3 from the outskirts (e.g. \citealt{AllenSchmidtFabian01}). As the X-ray surface brightness is proportional to the integral of the density squared along a line of sight, peaked surface brightness profiles imply that the gas in the cores is cooling much more rapidly than in the outskirts.  In the centre, the mean radiative cooling time, often calculated as the ratio of the luminosity of a region to its enthalpy, often drops below 1~Gyr.

Assuming steady-state and the absence of heating, there will be material with short cooling times in the cluster centre cooling out of the X-ray emitting band. As the volume of this material becomes much smaller, there must be a flow of material in to replace it to preserve the steady state. The steep surface brightness profiles imply rates of 10s to 1000s of solar masses per year cooling out of the X-ray band (see \citealt{Fabian94} for a review). This cooling material would be expected to eventually give rise to star formation.

This picture changed when high spectral resolution X-ray studies of nearby clusters of galaxies using the Reflection Grating Spectrometer (RGS) instruments on \emph{XMM-Newton} revealed a lack of cool X-ray emitting gas in these objects \citep{Tamura01b, Tamura01a, Peterson01, Kaastra01, Sakelliou02, Peterson03}. The main spectral indicator missing in these spectra are the emission lines of Fe~\textsc{xvii}, which are strong from material between 0.15 and 0.8~keV (1.7 to 9.4~MK).

The general accepted picture is that there is a lack of material below a factor 2 or 3 of the outer temperature, which matches the results from \emph{Chandra} spatially-resolved temperature profiles. This is not completely correct, however. Deep observations of nearby clusters show a much larger range of temperature. Fe~\textsc{xvii} emission lines have been seen in Centaurus \citep{SandersRGS08}, showing a temperature range of more than 10, Abell 2204 \citep{SandersA220409}, showing a range of 15, 2A~0335+096 \citep{Sanders2A033509}, showing a range of 8. Fe~\textsc{xvii} emission was also detected in Abell~262, M87 and NGC~533 in a large study by \cite{Peterson03}. There is much less material than would be expected to be seen in the case of steady-state radiative cooling without any heating, however.

There is evidence that the central active galactic nuclei (AGN) in these objects can energetically prevent much of the cooling (see reviews by \citealt{PetersonFabian06} and \citealt{McNamaraNulsen07}). This heating may be by the inflation of cavities (radio lobes) by the AGN or the subsequent dissipation of sound waves generated by the inflation. Such cavities are found in almost all nearby clusters which require heating \citep{DunnFabian06}.

If AGNs are responsible for preventing cooling in cluster cores, there are still remaining issues. In Centaurus the cooling time of the lowest detected component is only $10^7$~yr \citep{SandersRGS08}. If no cooling is taking place, then feedback must be able to operate on these timescales. Feedback must also be able to operate over the much longer cluster lifetime. Star formation in Centaurus must have occured slowly over the last 8~Gyr, or all must have been at earlier times \citep{SandersEnrich06}. 

We assume a Hubble constant of $70 \kmpspMpc$ and use the relative Solar metallicities of \cite{AndersGrevesse89}.

\begin{table*}
    \caption{Details of the targets and individual \emph{XMM-Newton} observations. The exposure times given are for the RGS1 instrument after cleaning. The absorption quoted is the Galactic absorption from \protect\cite{Kalberla05}. References in superscript for redshifts are (1) \protect\cite{StrubleRood99}, (2) \protect\cite{Johnstone98} and (3) \protect\cite{ZabludoffMulchaey00}.}
    \begin{tabular}{llllll}
    \hline
    Cluster      & Redshift & Absorption  & Observation & Date       & Exposure \\
                 &    & ($10^{20}\psqcm$) &             &            & (ks)\\
    \hline
    Abell 262    &$0.0163^1$& 5.67 & 0109980101  & 2001-01-16 & 26.2  \\
                 &          &      & 0504780101  & 2007-07-12 & 121.2 \\
                 &          &      & 0504780201  & 2007-07-18 & 41.4  \\
                 &          &      & total       &            & 188.8 \\
    Abell 3581   &$0.0218^2$& 4.36 & 0205990101  & 2004-01-29 & 43.5  \\
                 &          &      & 0504780301  & 2007-08-01 & 116.3 \\
                 &          &      & 0504780401  & 2007-08-03 & 27.7  \\
                 &          &      & total       &            & 187.5 \\
    HCG 62       &$0.01453^3$&3.31 & 0112270701  & 2003-01-15 & 12.4  \\
                 &          &      & 0504780501  & 2007-06-26 & 110.1 \\
                 &          &      & 0504780601  & 2007-06-29 & 33.6  \\
                 &          &      & total       &            & 156.6 \\
    \hline
    \end{tabular}
    \label{tab:sample}
\end{table*}

\begin{figure*}
    \includegraphics[width=0.8\textwidth]{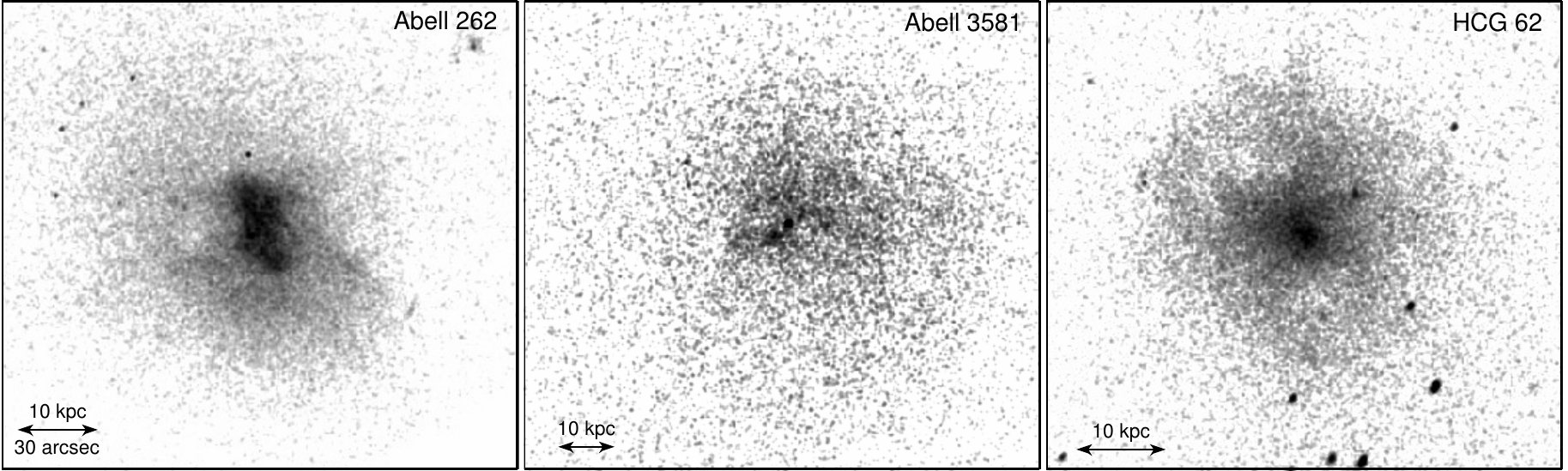}
    \caption{\emph{Chandra} images of the cores of the three objects examined in this paper. Images have been smoothed with a 1 arcsec Gaussian. These images show the \emph{Chandra} observations in Table~\ref{tab:chandraobs}.}
    \label{fig:chandraimages}
\end{figure*}

\begin{figure*}
    \centering
    \includegraphics[width=0.7\textwidth]{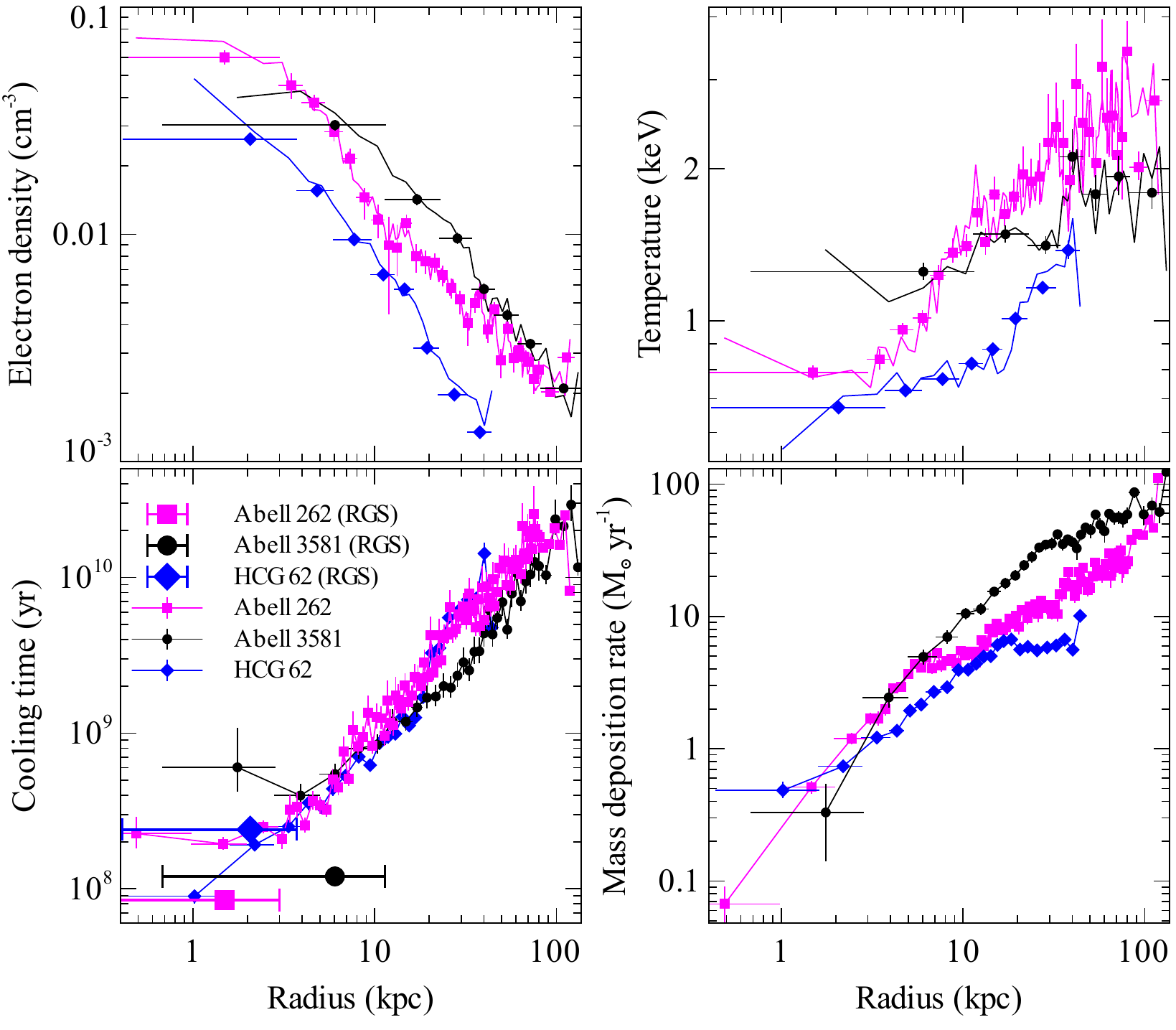}
    \caption{(Top left panel) Deprojected densities measured from \emph{Chandra} spectral fitting (shown as points) and from surface brightness deprojection (shown as solid lines). (Top right panel) Deprojected temperature measured from \emph{Chandra} spectral fitting (shown as points) and surface brightness deprojection (shown as lines). (Bottom left panel) Cooling time profiles measured from surface brightness deprojection (shown as small points) and RGS (shown as large points). (Bottom right panel) Cumulative mass deposition rates derived from \emph{Chandra} surface brightness deprojection.}
    \label{fig:chandraprofiles}
\end{figure*}

\begin{table}
    \caption{\emph{Chandra} datasets analysed.}
    \begin{tabular}{llll}
        \hline
        Target       & Obs-ID & Date       & Clean exposure (ks)\\
        \hline
        Abell 262    & 7921   & 2006-11-20 & 110.7 \\
        Abell 3581   & 1650   & 2001-06-07 & 7.2 \\
        HCG 62       & 921    & 2000-01-25 & 48.4 \\
        \hline
    \end{tabular}
    \label{tab:chandraobs}
\end{table}

\section{The sample}
We are investigating cool gas in a sample of three nearby objects with a range of mass scales. Table~\ref{tab:sample} shows the objects, redshifts, Galactic absorptions, \emph{XMM-Newton} observations, observation dates and exposures.

The \emph{Chandra} data of Abell~262 were examined by \cite{Blanton04} and \cite{Clarke09}. The cluster lies in the Perseus supercluster and has a bolometric X-ray luminosity of $4.4\times10^{43}\ergps$ from \emph{Einstein} \citep{David93}. Its cD galaxy is NGC~708. There is extensive interaction between the radio source and X-ray emitting gas. There is a tunnel like region of lower X-ray surface brightness to the south west which is coincident with an extension of the radio source. Towards the east there are three clumps of radio emission. The innermost one is coincident with an X-ray cavity. There is an additional cavity to the east. The bright X-ray structures are in the same regions as the optical line emission. \cite{Clarke09} put a timescale on the outburst repetition timescale of the central nucleus of 28~Myr. They claim that the total AGN output is sufficient to offset radiative cooling over several episodes of outbursts. The coolest X-ray gas seen by \emph{Chandra} is around 0.8~keV \citep{Blanton04}.

The short \emph{Chandra} observation of the galaxy cluster Abell~3581, which contains the radio source PKS~1404-267, was examined by \cite{Johnstone05}. They found that the X-ray temperature declines to around 0.4 of the maximum within the central 5 kpc and where the entropy drops to below $10 \keV\cmsq$. There are clear X-ray cavities located at the position of the radio source at 1.4~GHz. \cite{White97} give an X-ray luminosity of $4.2\times10^{43} \ergps$ for the cluster, correcting to our cosmology.

\cite{Morita06} examined \emph{Chandra} and \emph{XMM-Newton} data of the bright Hickson compact galaxy group HCG~62. They find two cavities in the X-ray emission. Two temperature components were required within the inner 2 arcmin (0.7 and 1.4~keV). The bolometric X-ray luminosity of the cluster is $5.6\times10^{42}\ergps$ \citep{PonmanBertram93}.

\section{Chandra data}
\emph{Chandra} observations of the clusters (listed in Table \ref{tab:chandraobs}) were analysed using \textsc{ciao} version 4.1.2 with \textsc{caldb} version 4.1.2 provided by the \emph{Chandra} X-ray Center (CXC).  The level 1 event files were reprocessed to apply the latest gain and charge transfer inefficiency correction and then filtered to remove bad grades.  The improved background screening provided by VFAINT mode was applied to the observations of Abell 262 and Abell 3581.  The background light curves of the level 2 event files were then filtered using the \textsc{lc\_clean} script\footnote{See http://cxc.harvard.edu/contrib/maxim/acisbg/} provided by M. Markevitch to remove periods affected by flares.  The final cleaned exposure times are shown in Table~\ref{tab:chandraobs}. Blank-sky background data sets available from the CXC were processed in the same way as the cluster observations and normalized to the source count rate in the $9-12\keV$ energy band.

Shown in Fig.~\ref{fig:chandraimages} are \emph{Chandra} images of each of the three objects.

\subsection{Deprojected quantities}
For each cluster, projected spectra were extracted in a series of concentric annuli centred on the emission peak and then deprojected with \textsc{dsdeproj} \citep{SandersPer07,Russell08} to remove the contributions from the outer cluster layers. We assume that the clusters are spherically symmetric in this analysis. The radial bins were chosen to ensure a minimum of 3\,000 deprojected counts in each deprojected spectrum.  Point sources were identified using the CIAO algorithm \textsc{wavdetect}, visually confirmed and excluded from the analysis.  All spectra were analysed in the energy range $0.5-7\keV$ and grouped with a minimum of 50 counts per spectral bin.  Response and ancillary response files were generated for each cluster spectrum, weighted according to the number of counts between $0.5$ and $7\keV$.  The deprojected spectra were fitted in \textsc{xspec} version 12.5.0 \citep{ArnaudXspec} with an absorbed thermal plasma emission model \textsc{phabs(mekal)} \citep{Balucinska92, Kaastra00, MeweMekal85, MeweMekal86, KaastraMekal92, LiedahlMekal95}. The absorbing column density was fixed to the Galactic values given by \cite{Kalberla05}, shown in Table \ref{tab:sample}.  The temperature, metallicity and model normalization were allowed to vary. We show deprojected temperature and electron density profiles in Fig.~\ref{fig:chandraprofiles}.

The lines shown on the temperature and density panels in Fig.~\ref{fig:chandraprofiles} are from a surface brightness deprojection (following \citealt{FabianPer81}) assuming an NFW potential \citep{NFW96}. We use a Metropolis-Hastings Markov Chain Monte Carlo (MCMC) method to iterate over the NFW parameters (concentration and $r_{200}$) and the outer pressure in the deprojection. The MCMC calculates the $\chi^2$ of the NFW-predicted gas temperature profile to the observed deprojected spectral temperature profile to determine whether to move to the new position in the chain. The parameters are stepped individually in random order in the MCMC using a Gaussian proposal probability distribution. The width of the proposal distribution is dynamically adjusted to achieve a repeat fraction of 0.75 during the burn in period of 10\,000 iterations. The lines show the median density and temperature profiles obtained with the MCMC. The input surface brightness profiles were created using radial bins a factor of three times smaller than the spectral bins for Abell~262 and HCG~62, and a factor of five for Abell~3581.

From the set of potentials, pressures, densities and temperatures obtained from the surface brightness deprojection, we can obtain mean radiative cooling time profiles and cooling flow mass deposition rates. These are shown in the bottom two panels of Fig.~\ref{fig:chandraprofiles}. The median values from the MCMC are shown with the 1-$\sigma$ errors calculated from the 15.9 and 84.2 percentiles. The mass deposition rates assume steady state and no heating and account for the gravitational contribution in the cooling.

\begin{table*}
\caption{Surface brightness deprojection results. The values shown are the median from the NFW MCMC analysis. The error bars were calculated from the 15.9 and 84.2 percentiles and are equivalent to 1-$\sigma$ errors. The values of $\sigma$ and $r_0$ were instead calculated using a nonsingular isothermal model (with the modified Hubble law form) in the MCMC analysis.}
\begin{tabular}{llllllll}
\hline
Cluster & Concentration & $r_{200}$ (Mpc) & $P_\mathrm{outer}$ (log$_{10}$ \ergpcmcu) &
$r_\mathrm{cool}$ (kpc) & $\dot{M}$ (\Msunpyr) & $\sigma$ (\kmps) & $r_{0}$ (\kpc) \\
\hline
Abell~262 & $31.6^{+1.6}_{-2.1}$ & $0.342^{+0.012}_{-0.008}$ & $-10.74 \pm 0.01$ &
39 & $18 \pm 1$ & $256 \pm 3$ & $4.7 \pm 0.3$
\\
Abell~3581 & $15.0^{+1.9}_{-2.4}$ & $0.57^{+0.06}_{-0.03}$ & $-11.04 \pm 0.04$ &
57 & $49 \pm 6$ & $323 \pm 7$ & $13.1 \pm 1.6$
\\
HCG~62 & $45.0^{+3.1}_{-2.7}$ & $0.271 \pm 0.008$ & $-11.19 \pm 0.02$ &
36 & $6.7 \pm 0.2$ & $222 \pm 3$ & $3.1 \pm 0.2$
\\
\hline
\end{tabular}
\label{tab:deproj}
\end{table*}

Shown in Table~\ref{tab:deproj} are the best fitting NFW parameters and outer pressure in the deprojection. We also show the cooling radius, where the cooling time is approximately the time since $z=1$. The listed mass deposition rate is the cumulative mass deposition rate inside this radius.

Note that the best fitting NFW concentrations are very high. This is likely to be because the \emph{Chandra} observations concentrate on the region where the potential of the central cluster galaxy is most important. Only a very small cluster volume is being sampled. We also tried using a nonsingular isothermal model with the modified Hubble law density approximation. The resulting density, temperature, cooling time and mass deposition rate profiles are almost identical with those computed using the NFW mass model. The median velocity dispersion, $\sigma$ and King radius, $r_0$, for the isothermal mass model are also shown in Table~\ref{tab:deproj}.

\begin{figure}
    \centering
    \includegraphics[width=0.8\columnwidth]{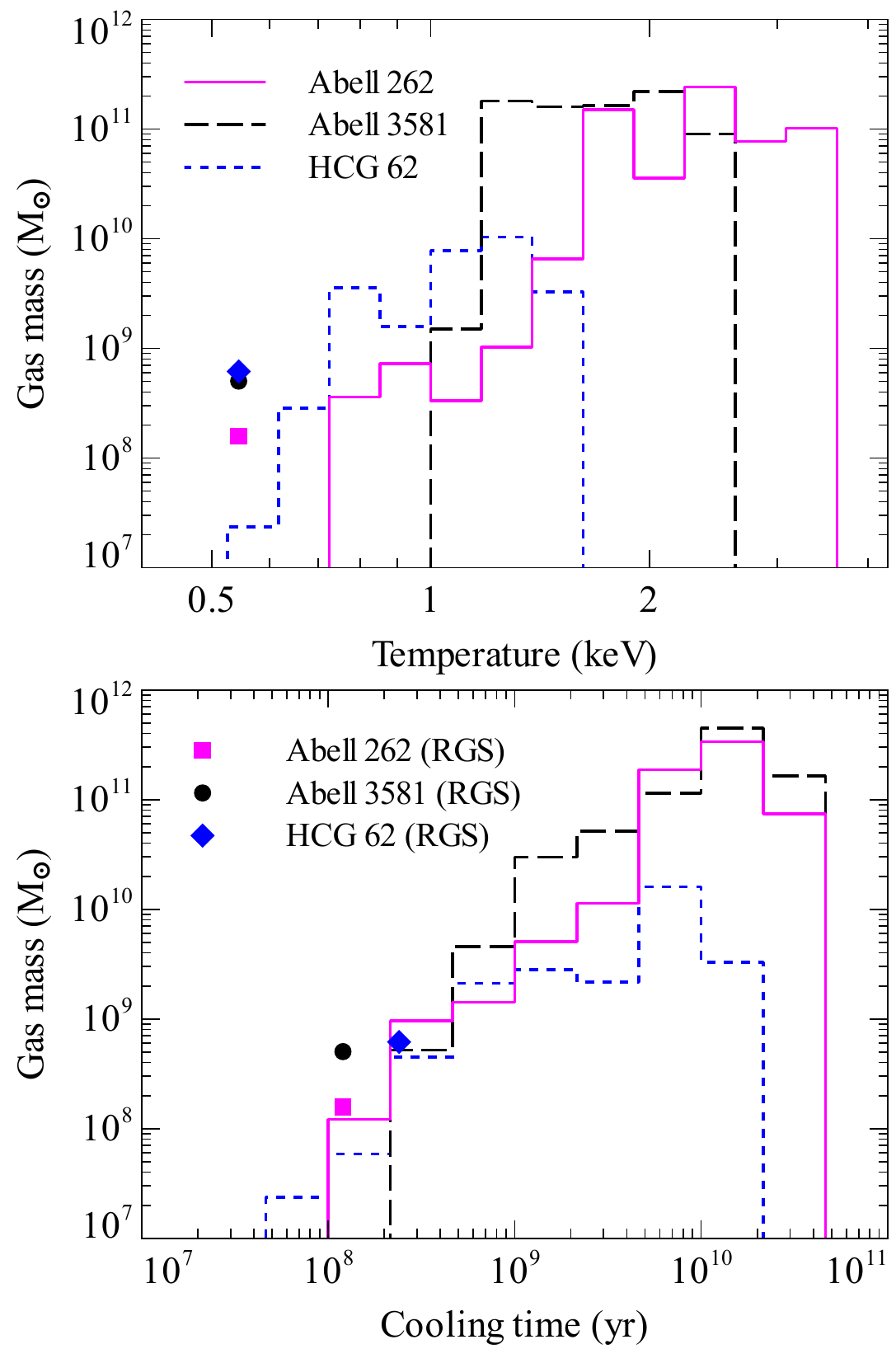}
    \caption{Distribution of gas mas as a function of mean radiative cooling time and temperature. The distributions are calculated from the \emph{Chandra} deprojection analysis (Fig.~\ref{fig:chandraprofiles}). The points show values calculated from the RGS emission measures of the 0.544 keV component, assuming that they are in pressure equilibrium with the gas from the \emph{Chandra} deprojection.}
    \label{fig:mass_t}
\end{figure}

Using these profiles we can calculate the distribution of gas mass as a function of temperature or cooling time. These are plotted in Fig.~\ref{fig:mass_t}.

\subsection{Temperature maps}
\begin{figure}
    \centering
    \includegraphics[width=\columnwidth]{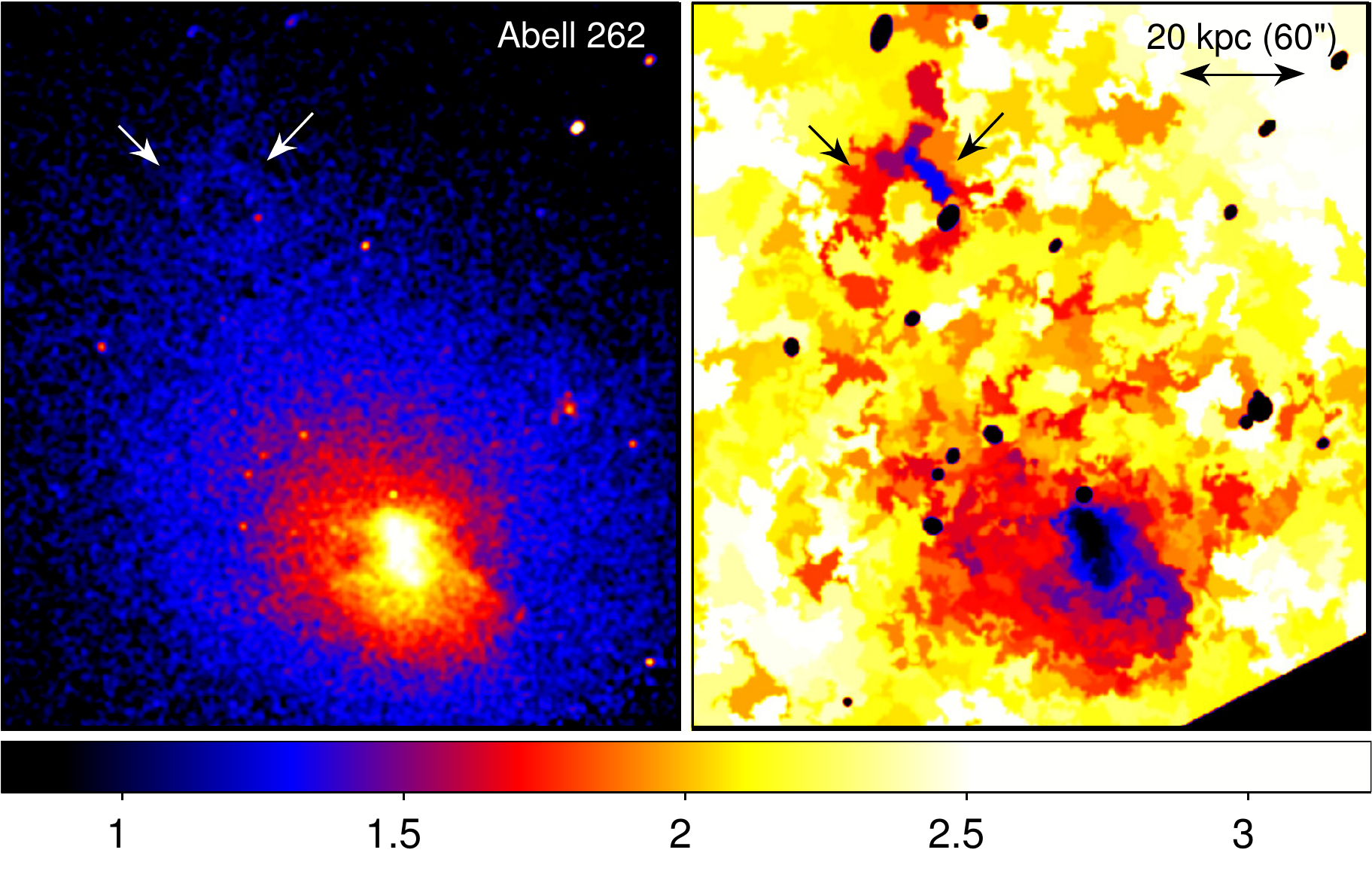}
    \includegraphics[width=\columnwidth]{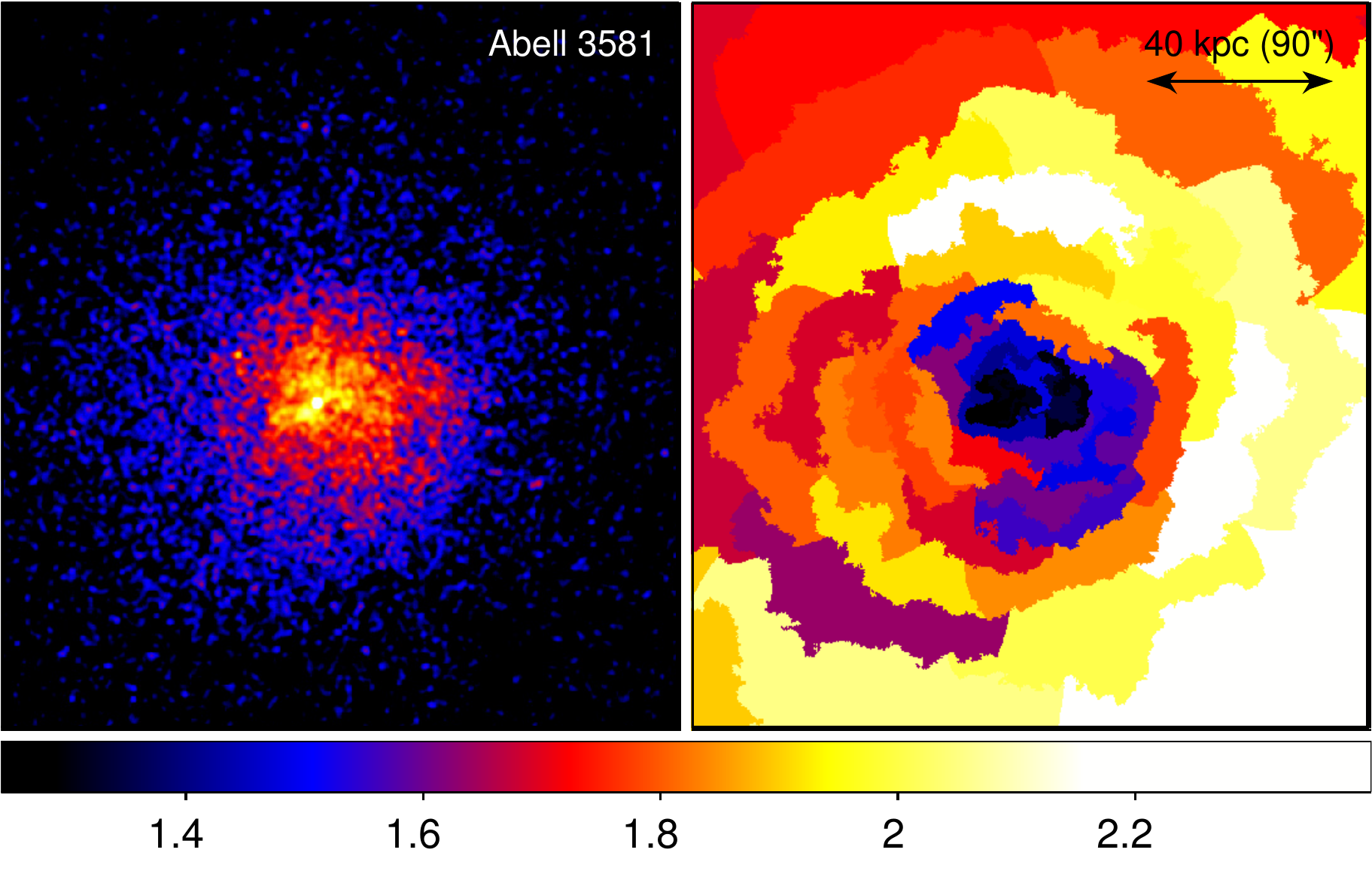}
    \includegraphics[width=\columnwidth]{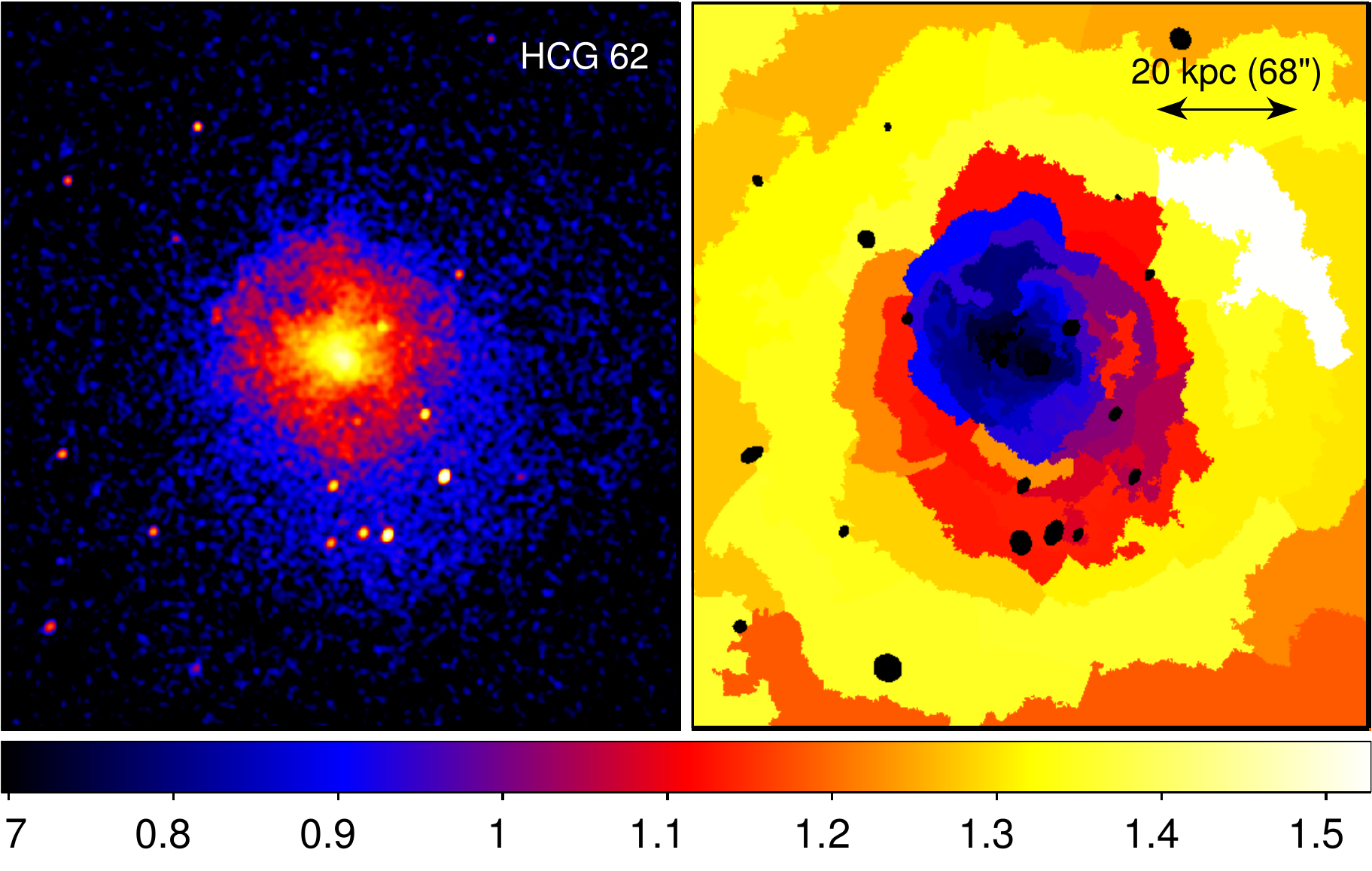}
    \caption{Projected emission-weighted temperature maps of each object from \emph{Chandra} data (right panels) next to X-ray images (left panel). The colour bars at the bottom of each pair of images show the temperature in keV. Note the ring of cool gas in Abell~262 to the north east (marked by arrows). Each region in the temperature map was chosen to have a signal to noise ratio of 22 (approximately 480 counts).}
    \label{fig:tmaps}
\end{figure}

We have calculated projected emission-weighted temperature maps of each of the clusters from the \emph{Chandra} data (Fig.~\ref{fig:tmaps}). Regions were selected to have a signal-to-noise ratio of 22 (corresponding to 480 counts in the absence of background) using the Contour Binning algorithm \citep{SandersBin06}. The regions were chosen to have a maximum ratio of the 2 between the longest and shortest dimensions. The spectra from the regions were fit between 0.5 and 7~keV with a single component \textsc{apec} spectral model \citep{SmithApec01} with free temperature, metallicity and normalization, minimizing the C-statistic in the fits. The Galactic absorption and redshift were fixed to the values given in Table~\ref{tab:sample}.

The temperature maps show a significant amount of structure. One interesting feature in Abell 262 is a 20 arcsec (6.7 kpc) radius ring of cool gas 3 arcmin (60 kpc) to the north east of the cluster centre. It is marked by arrows in the top panels of Fig.~\ref{fig:tmaps}. The ring appears that it may be connected to the centre of the cluster with a cool ridge and the cool gas may extend northwards beyond the ring. The ring is not associated with any galaxies.

\section{RGS data analysis}
\label{sect:rgsanalysis}
We extracted spectra from each observation listed in Table~\ref{tab:sample} using \textsc{rgsproc}, part of \textsc{xmmsas} (version 8.0.0). For the main part of the analysis we extracted spectra from 95 per cent of the cross-dispersion point-spread-function (PSF) and within 90 per cent of the pulse-height distribution. The PSF extraction region corresponds to a roughly 100 arcsec strip across the cluster centre. As the cluster emission fills the RGS field of view, we generated template background spectra for each observation with \textsc{rgsbkgmodel}. Each of the observations for a target were processed using `attstyle' option in \textsc{rgsproc} set to `user' so that the spectra could be added together using \textsc{rgscombine}. As the wavelengths in the spectrum are dependent on the correct position of the source, we examined the \emph{Chandra} observations by eye to find the coordinates of the peak diffuse X-ray emission.

\begin{figure*}
  \includegraphics[width=0.8\textwidth]{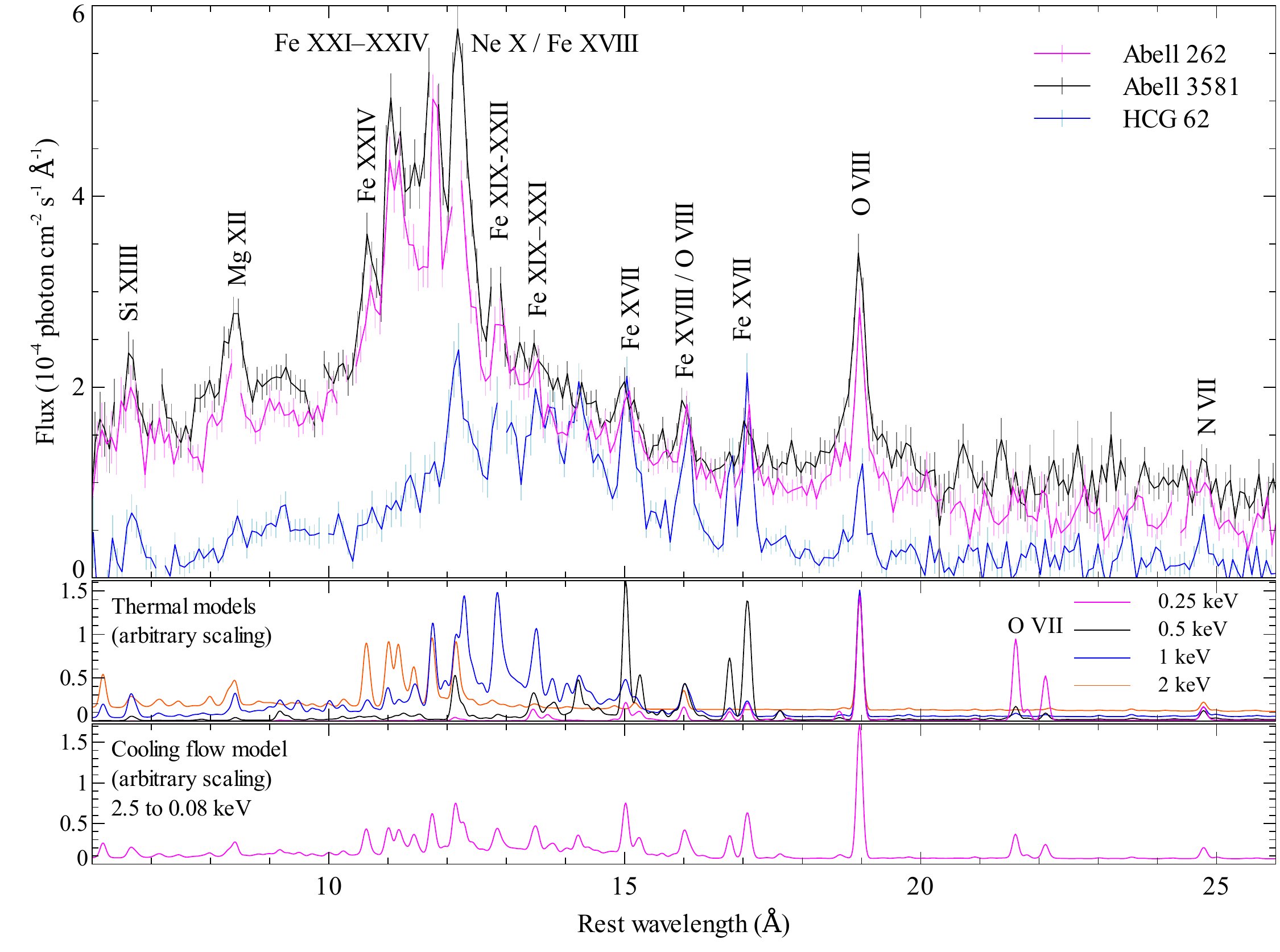}
  \caption{(Top panel) The fluxed spectra as a function of rest wavelength. The stronger emission lines have been labelled. (Central panel) Model solar metallicity spectra at four different temperatures for comparison, broadened to comparable spectral resolution. (Bottom panel) Spectrum of a solar metallicity cooling flow model cooling from 2.5 to 0.0808 keV.}
  \label{fig:fluxedspecs}
\end{figure*}

\begin{figure}
    \includegraphics[width=\columnwidth]{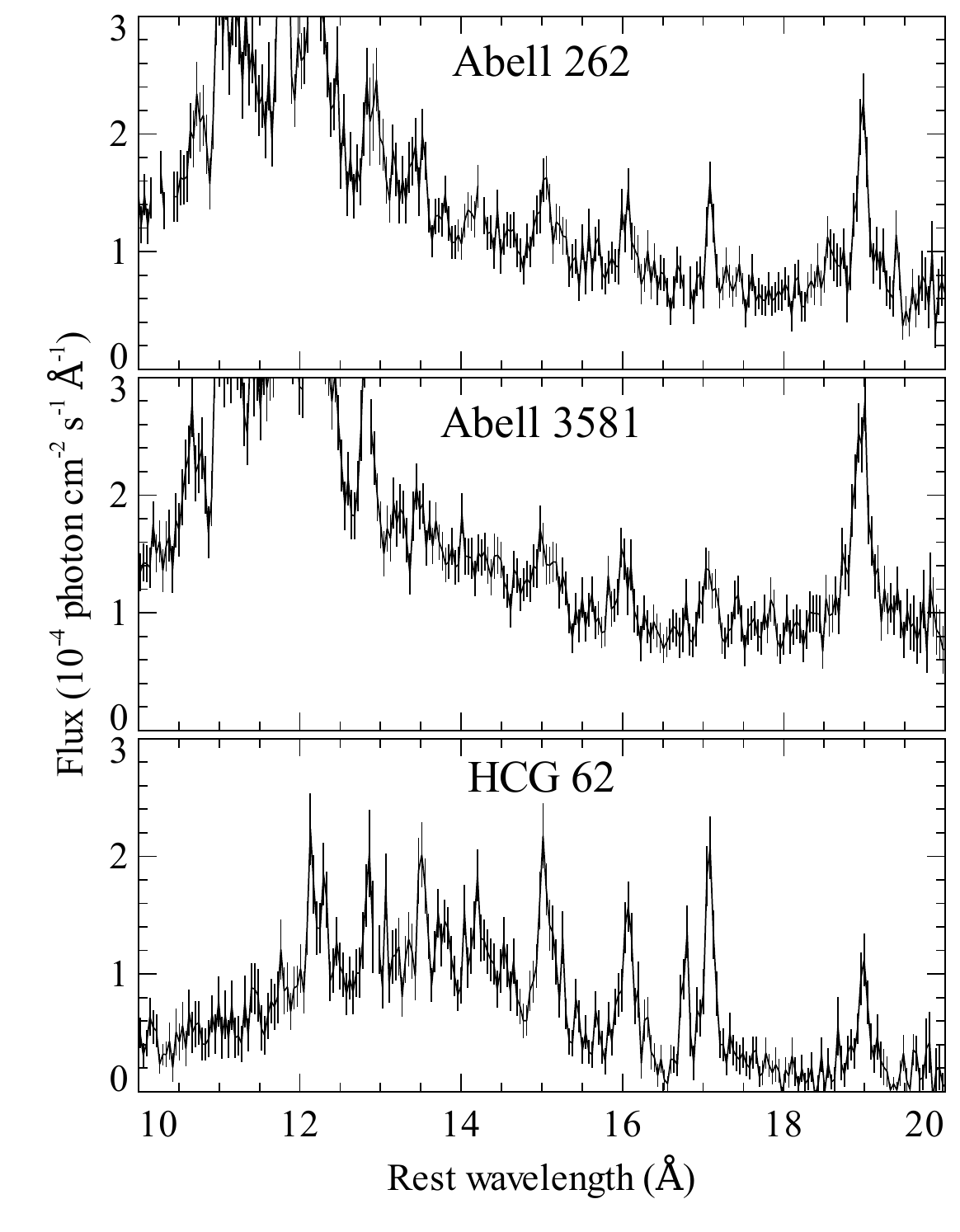}
    \caption{Fluxed spectra, extracted from 90\% extraction region and corrected to rest wavelength.}
    \label{fig:fluxedspecs090}
\end{figure}

We applied time filters to remove the strongest flares. Although there are periods remaining in the observations containing relatively mild flares, the template background generator is able to generate appropriate backgrounds by adding together backgrounds with similar count off-axis count rates. After background subtraction, each observation of each cluster matches the other observations well at long wavelengths, where background is most important. We use a 7 to 26{\AA} wavelength range to minimize background effects.

The observations for each cluster were combined with the \textsc{rgscombine} tool. The template backgrounds were combined with our own program which averaged them together, weighting by the exposure time of the respective foreground spectrum.

Combined fluxed spectra from each of the targets, produced by \textsc{rgsfluxer} and removing model backgrounds are shown in  Fig.~\ref{fig:fluxedspecs}. In the lower panel we show similarly broadened model spectra at four different temperatures for comparison.  Although the fluxed spectra cannot be used for spectral analysis, they show that the spectra contain Fe~\textsc{xvii} emission lines indicating material around 0.7~keV (the ionisation fraction of Fe in Fe~\textsc{xvii} peaks at 0.35~keV; \citealt{Mazzotta98}). They do not, however, show evidence for O~\textsc{vii} lines, indicative of gas below 0.25~keV (O~\textsc{vii} ion fraction peaks at 0.07~keV).

The spectral features from the innermost region can be seen more clearly from spectra extracted using a smaller extraction region. Shown in Fig.~\ref{fig:fluxedspecs090} are the spectra extracted from 90 per cent of the PSF, which corresponds to roughly a 50~arcsec width strip across the cluster.

\subsection{Spectral fitting}
\label{sect:specfitting}
The spectral fitting procedure here broadly follows that used by \cite{SandersRGS08}.  Given the wavelength range of the RGS spectra where the background in low, we allow the N, O, Ne, Mg, Ca, Fe and Ni metallicities to vary independently. We fix the metallicities of C, Si, S and Ar to the Fe metallicity, as we do not have good constraints for these elements. We use the \textsc{phabs} absorption model, fixing the absorption column density to the Galactic value (as obtained from \citealt{Kalberla05}; shown in Table \ref{tab:sample}). We only fit the first order spectra in this analysis, using the wavelength range between 7 and {26\AA}. The relative normalization of the RGS2 instrument was allowed to vary relative to RGS1. We grouped the spectra to have a minimum of 20 counts per spectral bin and minimized the $\chi^2$ in the spectral fit. We did not use the C statistic here because the template background spectra are synthetic count rate spectra, not count spectra, so Poisson statistics cannot be assumed.

As clusters are extended objects, we need to account for the spectral broadening due to the spatial extent. This is done using a smoothing model \citep{SandersRGS08}. We use an upper limit of 3 arcmin for the broadening when fitting. Different broadenings are used for components where feasible. We caution that the metallicities from the spectral fitting may be low because as we assume a Gaussian broadening from the spatial extent and do not fully model the wings of the emission lines. Metallicity ratios are more likely to be accurate than the absolute values.

It is possible to use the spatial information from imaging with \emph{Chandra} to account for the spatial broadening of the emission lines, using for example the \textsc{rgsxsrc} model in \textsc{xspec}. This approach assumes that the spectral lines are emitted from the same region as broad band imaging and that the X-ray spectrum is uniform over this region. This is likely to be an incorrect assumption if the cooler gas is more centrally concentrated as suggested by the temperature profiles and maps (Fig.~\ref{fig:chandraprofiles} and Fig.~\ref{fig:tmaps}). We therefore use separate free parameters for the spatial broadening for the hotter and cooler components. Our spectral fitting of the RGS data confirms that the cooler gas is more spatially concentrated. The good agreement between the simple spectral fitting used here and the smooth particle analysis in Section \ref{sect:smoothparticle}, which treats the spatial extent of the source self consistently, implies the simple broadening approach works well.

\begin{table*}
    \caption{Spectral fitting results. (1) Temperatures of the two components for $2\times$\textsc{vapec}, upper temperature of \textsc{cevmkl} and \textsc{vapec} temperature for \textsc{vapec+cevmkl}, upper temperature of \textsc{mkcflow} and \textsc{vapec} temperature for \textsc{vapec+mkcflow} and lower temperature of \textsc{mkcflow} if there is a free minimum. (2) Emission measures of thermal or \textsc{cevmkl} components multiplied by $10^4$. (3) Mass deposition rates in $\Msunpyr$ for \textsc{mkcflow} components or (4) emissivity index $\alpha$ for \textsc{cevmkl} components. (5) Smoothing scales in arcmin for first and second components, respectively. The N, O, Ne, Mg, Ca, Fe and Ni columns show metallicities in Solar units. Other metallicities are tied to Fe, except for He.}
\begin{tabular}{cccccccccc}
\hline
Model & Cluster & $\chi^2_\nu$ & $kT_1$ (1) &$\mathcal{E}_1$ (2)&$\dot{M}$ (3)&size$_1$ (5)& N  & O  & Ne \\
      &         &              & $kT_2$ (1) &$\mathcal{E}_2$ (2)& $\alpha$ (4)&size$_2$ (5)& Mg & Ca & Fe \\
      &         &              &            &                   &             &            & Ni \\
\hline
$2\times$\textsc{vapec}& A262 & $1.20=$     & $0.72 \pm 0.02$ & $6.9 \pm 0.4$  & & $0.34 \pm 0.06$ & $0.23 \pm 0.17$ & $0.25 \pm 0.02$ & $0.27 \pm 0.07$  \\
                      &       & $1488/1238$ & $1.49 \pm 0.03$ & $65.2 \pm 1.5$ & & $0.73 \pm 0.06$ & $0.40 \pm 0.06$ & $2.6 \pm 0.5$   & $0.37 \pm 0.03$  \\
                      &       &             &                 &                & &                 & $1.47 \pm 0.15$ \\
                      & A3581 & $1.14=$     & $0.69 \pm 0.03$ & $4.3 \pm 0.5$&&$0.54^{+0.17}_{-0.13}$& $0.24\pm0.13$ & $0.24\pm0.01$  & $0.31 \pm 0.06$ \\
                      &       & $1674/1466$ & $1.41 \pm 0.02$ & $89 \pm 2$     & & $0.71 \pm 0.04$ & $0.43 \pm 0.05$ & $2.41 \pm 0.40$ & $0.32 \pm 0.02$ \\
                      &       &             &                 &                & &                 & $0.82 \pm 0.11$ \\
                      & HCG62 & $1.18=$     & $0.69 \pm 0.01$ & $10.0\pm 0.9$  & & $0.34 \pm 0.03$ & $0.54 \pm 0.21$ & $0.228^{+0.036}_{-0.028}$ & $0.44 \pm 0.09$\\
                      &       & $653.8/553$ & $1.07 \pm 0.04$ & $9.0 \pm 0.9$ &&$1.6^{+0.9}_{-0.4}$& $0.58 \pm 0.09$ & $0.7 \pm 0.5$ & $0.43 \pm 0.04$ \\
                      &       &             &                 &                & &                 & $0.8^{+0.3}_{-0.2}$ \\
\hline
\textsc{vapec}+       & A262  & $1.21=$     & $1.63 \pm 0.06$ & $37 \pm 4$  &               &$1.7 \pm 0.3$   & $0.23 \pm 0.18$ & $0.28 \pm 0.03$ & $0.28 \pm 0.06$ \\
\textsc{cevmkl}       &       & $1493/1238$ &                 & $150 \pm 23$&$2.56 \pm 0.16$&$0.41 \pm 0.04$ & $0.44 \pm 0.08$ & $2.9 \pm 0.6$   & $0.42 \pm 0.04$ \\
                      &       &             &                 &             &               &                & $1.54 \pm 0.18$ \\
                      & A3581 & $1.14=$     & $1.49 \pm 0.04$ & $65 \pm 1$  &               &$0.9 \pm 0.2$   & $0.26 \pm 0.14$ & $0.25 \pm 0.02$ & $0.34 \pm 0.06$ \\
                      &       & $1670/1466$ &             &$140^{+70}_{-50}$&$3.1 \pm 0.5$  &$0.45 \pm 0.08$ & $0.43 \pm 0.06$ & $2.55 \pm 0.44$ & $0.35 \pm 0.02$ \\
                      &       &             &                 &             &               &                & $0.77^{+0.13}_{-0.06}$\\
                      & HCG62 & $1.28=$     & $0.87 \pm 0.01$ &$5.7 \pm 0.9$&               &$>2.7$          & $0.33 \pm 0.13$ & $0.15 \pm 0.02$ & $0.42 \pm 0.06$ \\
                      &       & $705.8/553$ &                 &$81 \pm 9$   &$3.1 \pm 0.3$  &$0.33 \pm 0.03$ & $0.56 \pm 0.08$ & $<0.56$         & $0.33 \pm 0.03$ \\
                      &       &             &                 &             &               &                & $1.1 \pm 0.2$ \\
\hline
\textsc{vapec}+       & A262  & $1.24=$     & $1.44 \pm 0.03$ & $61 \pm 2$  &$4.4 \pm 0.3$  &$0.77 \pm 0.09$ & $0.12 \pm 0.11$ & $0.19 \pm 0.01$ & $0.23 \pm 0.06$ \\
\textsc{mkcflow}      &       & $1534/1239$ &                 &             &               &$0.43 \pm 0.06$ & $0.41 \pm 0.06$ &  $2.4 \pm 0.5$  & $0.32 \pm 0.03$ \\
                      &       &             &                 &             &               &                & $1.68 \pm 0.15$ \\
                      & A3581 & $1.15=$     & $1.40 \pm 0.02$ & $87 \pm 2$  &$4.6 \pm 0.5$  &$0.71 \pm 0.05$ & $0.18 \pm 0.11$ & $0.21 \pm 0.01$ & $0.29 \pm 0.05$ \\
                      &       & $1683/1467$ &                 &             &         &$0.56^{+0.18}_{-0.13}$& $0.43 \pm 0.05$ & $2.4 \pm 0.4$ & $0.31 \pm 0.02$ \\
                      &       &             &                 &             &               &                & $0.88 \pm 0.12$ \\
                      & HCG62 & $1.35=$     & $0.83 \pm 0.02$ & $16 \pm 1$  &$3.6 \pm 0.3$  &$0.76 \pm 0.12$ & $0.20 \pm 0.09$ & $0.10 \pm 0.01$ & $0.36 \pm 0.06$ \\
                      &       & $746.1/554$ &                 &             &               &$0.19 \pm 0.05$ & $0.50 \pm 0.07$ & $<0.27$         & $0.29 \pm 0.02$ \\
                      &       &             &                 &             &               &                & $1.23 \pm 0.16$ \\
\hline
\textsc{vapec}+       & A262  & $1.20=$     & $1.59 \pm 0.05$ & $53 \pm 2$  &$6.9 \pm 0.5$&$0.9^{+0.2}_{-0.1}$&$0.25 \pm 0.19$ & $0.27 \pm 0.03$ & $0.28 \pm 0.07$ \\
\textsc{mkcflow}      &       & $1489/1238$ & $0.53 \pm 0.02$ &             &               &$0.34 \pm 0.05$ & $0.41 \pm 0.07$ & $2.8 \pm 0.6$   & $0.40 \pm 0.04$ \\
(free minimum)        &       &             &                 &             &               &                & $1.51 \pm 0.17$ \\
                      & A3581 & $1.14=$     & $1.45 \pm 0.03$ & $82 \pm 3$  &$8 \pm 1$      &$0.74 \pm 0.06$ & $0.25 \pm 0.14$ & $0.25 \pm 0.02$ & $0.32 \pm 0.06$ \\
                      &       & $1672/1466$ & $0.52 \pm 0.05$ &             &         &$0.49^{+0.16}_{-0.11}$& $0.43 \pm 0.05$ & $2.5 \pm 0.4$   & $0.33 \pm 0.02$ \\
                      &       &             &                 &             &               &                & $0.80 \pm 0.12$ \\
                      & HCG62 & $1.22=$     & $0.96 \pm 0.02$ & $6.7 \pm 1$ &$10.0 \pm 0.9$ & $>2.6$         & $0.45 \pm 0.19$ & $0.20 \pm 0.02$ & $0.42 \pm 0.06$ \\
                      &       & $675.4/553$ & $0.51 \pm 0.02$ &             &               & $0.34 \pm 0.03$& $0.57 \pm 0.07$ & $<0.86$         & $0.37^{+0.03}_{-0.02}$\\
                      &       &             &                 &             &               &                & $0.9^{+0.2}_{-0.1}$ \\
\hline
$6\times$\textsc{vapec}& A262 & $1.19=$     & See             & See         &         &$0.81^{+0.08}_{-0.07}$& $0.29 \pm 0.19$ & $0.28 \pm 0.03$ & $0.32^{+0.10}_{-0.07}$\\
                      &       & $1464/1235$ &Fig.~\ref{fig:6cmpt_em}&Fig.~\ref{fig:6cmpt_em}&&$0.33 \pm 0.06$ & $0.40 \pm 0.08$ & $2.9 \pm 0.6$   & $0.48^{+0.06}_{-0.04}$\\
                      &       &             &                 &             &               &                & $1.59 \pm 0.19$ \\
                      & A3581 & $1.10=$     &                 &             &               &$0.72 \pm 0.05$ & $0.29 \pm 0.14$ & $0.26 \pm 0.02$ & $0.34 \pm 0.07$ \\
                      &       & $1602/1463$ &                 &             &               &$0.57 \pm 0.16$ & $0.40 \pm 0.06$ & $2.4 \pm 0.5$   & $0.38 \pm 0.03$ \\
                      &       &             &                 &             &               &                & $0.95 \pm 0.15$ & \\
                      & HCG62 & $1.39=$     &                 &             &               &$1.7 \pm 0.7$   & $0.42 \pm 0.18$ & $0.20 \pm 0.02$ & $0.41 \pm 0.07$ \\
                      &       & $764/550$   &                 &             &               &$0.44 \pm 0.03$ & $0.49 \pm 0.08$ & $<0.99$         & $0.44^{+0.04}_{-0.06}$\\
                      &       &             &                 &             &               &         & $1.04^{+0.15}_{-0.23}$ & \\
\hline
$6\times$\textsc{vmcflow}&A262& $1.19=$     & See             & See         &         &$0.77^{+0.08}_{-0.06}$& $0.31 \pm 0.19$ & $0.28 \pm 0.02$ & $0.31^{+0.09}_{-0.07}$\\
\textsc{(apec)}       &       & $1467/1235$ &Fig.~\ref{fig:vmcflow_apec_spex}&Fig.~\ref{fig:vmcflow_apec_spex}&
                                                                                            & $0.36 \pm 0.06$& $0.40 \pm 0.08$ & $3.0 \pm 0.6$   & $0.47^{+0.04}_{-0.02}$\\
                      &       &             &                 &             &               &                & $1.69 \pm 0.19$ \\
                      & A3581 & $1.10=$     &                 &             &               & $0.73 \pm 0.05$& $0.33^{+0.08}_{-0.14}$&$0.27\pm0.02$&$0.41 \pm 0.06$\\
                      &       & $1609/1463$ &                 &             &               & $0.49 \pm 0.16$& $0.41 \pm 0.06$ & $2.8 \pm 0.5$   & $0.44 \pm 0.02$ \\
                      &       &             &                 &             &               &                & $0.88 \pm 0.15$ \\
                      & HCG62 & $1.36=$     &                 &             &               & $>1.9$         & $0.53 \pm 0.02$ & $0.24^{+0.02}_{-0.04}$&$0.49 \pm 0.09$\\
                      &       & $747/550$   &                 &             &               & $0.43 \pm 0.03$& $0.62 \pm 0.09$ & $0.7 \pm 0.6$   & $0.48^{+0.03}_{-0.07}$\\
                      &       &             &                 &             &               &                & $1.1^{+0.3}_{-0.2}$\\
\hline
\end{tabular}
\label{table:fitresults}
\end{table*}

\begin{figure}
    \includegraphics[width=\columnwidth]{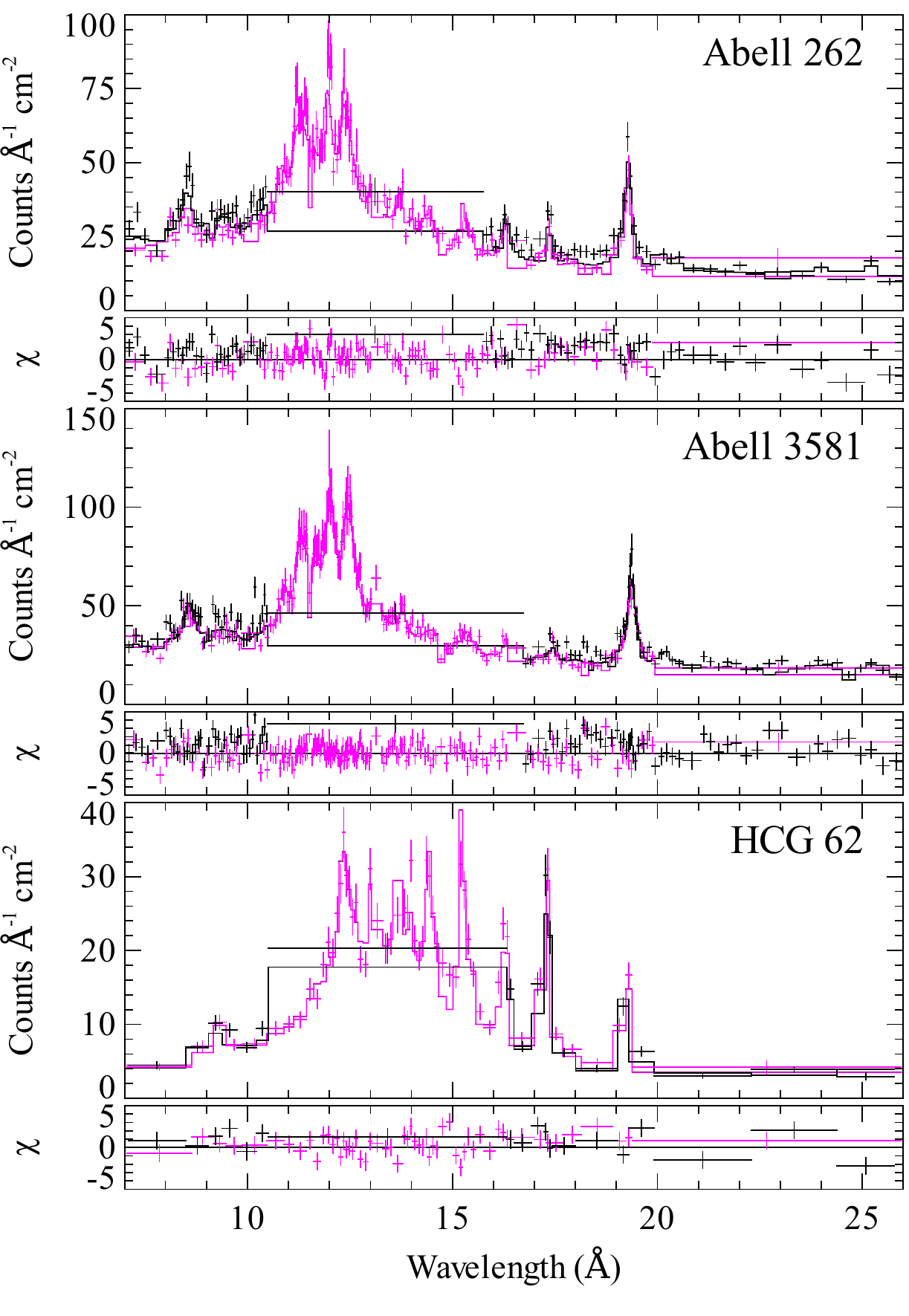}
    \caption{Two component \textsc{vapec} fits to the RGS data extracted from each object. The models are the solid lines. The data are shown as points. The spectra and models from the two RGS detectors are plotted in black and magenta. Some spectral regions are covered by only one detector. The panels with the axis scale $\chi$ show the residuals of the fits. The data and models are shown dividing through by the effective area of the instrument as a function of wavelength. The data have been rebinned to have a signal to noise ratio of at least 10 in each spectral bin.}
    \label{fig:2temp}
\end{figure}

\subsubsection{Two temperature fits}
We first examined some fairly simple spectral models. As a single temperature model is clearly insufficient to fit the data we used a model consisting of two \textsc{apec} components with the variable elemental abundances listed above (we refer to this model as $2\times$\textsc{vapec}). Each of the thermal components were allowed to have a different spatial broadening and the metallicities of the components were tied to be the same. The temperatures and model normalizations were allowed to be free. The best fitting parameter values and uncertainties are shown in Table~\ref{table:fitresults}. The spectra with the best fitting models are shown in Fig.~\ref{fig:2temp} with the residuals from the fit.

In \textsc{xspec} the normalization of the thermal components (proportional to the emission measure) is defined as
\begin{equation}
\mathcal{E} = \frac{10^{-14}}{4\pi \left[D_A(1+z)\right]^2}
\int n_\mathrm{e}\,n_\mathrm{H}\,\mathrm{d}V,
\end{equation}
where $D_A$ is the angular diameter distance to the object in cm, $V$ is the volume in the cluster being examined in $\cmcu$, $z$ is the redshift and $n_\mathrm{e}$ and ${n_\mathrm{H}}$ are the electron and Hydrogen number densities, respectively, in $\pcmcu$. The units of $\mathcal{E}$ are cm$^{-5}$.

The two temperature models are reasonable fits to the data, given the systematic issues with fitting high spectral resolution data. The lower temperature value for the three targets is close to 0.7~keV. This component is also broadened less than the hotter components, being consistent with it being emitted from a smaller region.

\subsubsection{Power law emissivity model}
Another simple parametrization is to have a power law distribution of emission measures as a function of temperature. The emission measure of each temperature component is proportional to $(T/T_\mathrm{max})^\alpha$, where $T_\mathrm{max}$ is a parameter for the upper temperature. This model is not physically motivated. We chose to have the spectra in the model constructed from \textsc{apec} spectra and also included another \textsc{apec} component to account for any non-cool gas.

We show the best fitting parameters for this model (which we refer to as \textsc{vapec+cevmkl}) in Table~\ref{table:fitresults}. In this model we find the best fitting indices for the power law distribution of 3.1 for Abell~3581 and HCG~62 and 2.6 for Abell~262. It is unclear whether this model is a useful parametrization as the relative balance between the \textsc{apec} and power-law emissivity models are quite different between the clusters. The \textsc{wdem} power-law emissivity model \citep{Kaastra04} with high and low temperature cut-offs appears to be a good parametrization for many clusters, though we do not investigate that model here, because our later analysis shows that the temperature distribution in these cool objects does not appear to be a powerlaw over a wide temperature range.

\subsubsection{Cooling flow models}
\label{sect:cflowmodels}
The spectrum can be fit by a cooling flow spectral model made up of a range thermal components where the emission measure distribution as a function of temperature is that of a gas which is radiatively cooling in a steady state and with no heating. Here we use a \textsc{vmkcflow} model to calculate this. We use the \textsc{apec} model to calculate its thermal components.

We first fitted a model where the gas is cooling to the minimum temperature (0.0808 keV). The upper temperature, mass deposition rate and metallicities are free parameters. We also include the non cooling gas as an \textsc{apec} component setting its temperature to the upper temperature of the \textsc{vmkcflow} model and using the same metallicities. We also allow for different spatial broadening for the thermal and cooling flow components.

The best fitting parameters for this spectral model are shown in  Table~\ref{table:fitresults}, listed under \textsc{vapec+mkcflow}. The mass deposition rates for the three objects are 4.4, 4.6 and $3.3\Msunpyr$ for Abell~262, Abell~3581 and HCG~62, respectively, where the cooling flow minimum temperature is fixed to zero.

If we instead fit a cooling model where the minimum temperature is allowed to be non-zero, we obtain mass cooling rates of 6.9, 9 and $10\Msunpyr$. The systems show very similar minimum temperatures of 0.53, 0.52 and 0.51~keV, for Abell~262, Abell~3581 and HCG~62, respectively.

\subsubsection{Multi-component thermal models}
\label{sect:multicomp}
To investigate more thoroughly the amount of gas as a function of temperature, we have fitted a model made up six thermal components at fixed temperature. We use temperatures of 0.272, 0.544, 0.862, 1.366, 1.719 and 2.164~keV. These particular temperatures were chosen because they are temperatures the \textsc{apec} model is calculated at in the table in \textsc{xspec}, and they are sufficiently separated in temperature in order to be resolved. The three coolest components were smoothed by a different spatial scale to the three hottest. We tried two different thermal spectral models, \textsc{apec} and the \textsc{mekal} model taken from \textsc{spex} version 2.00.11. With this model we fix the temperature but allow the normalizations to vary to examine the emission measure distribution.

\begin{figure*}
    \includegraphics[width=0.7\textwidth]{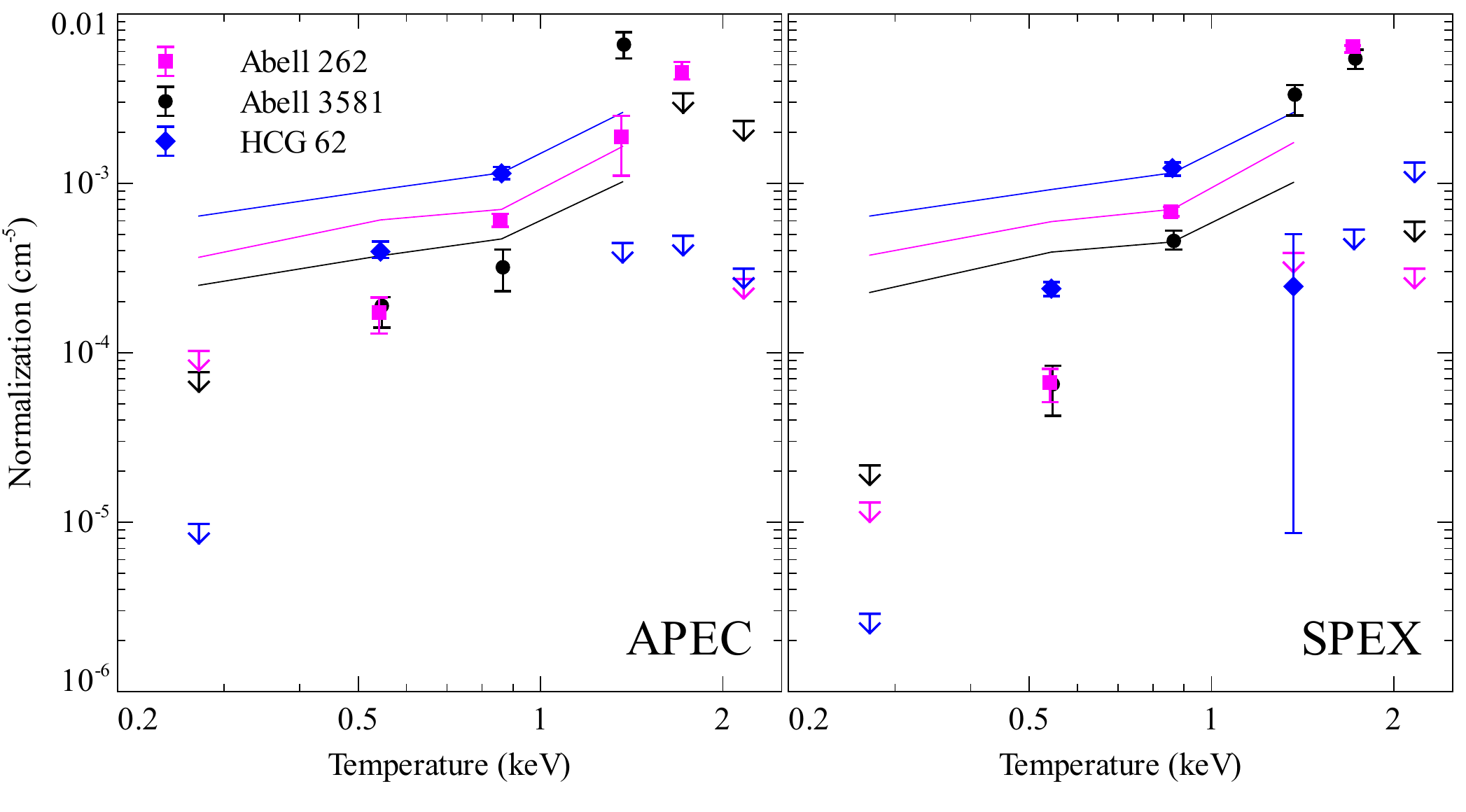}
    \caption{Normalizations for the temperature components in a 6 component spectral model. The left and right panels show the results for \textsc{apec} and \textsc{spex} models, respectively. The solid lines show the expected distribution of normalization for a cooling flow of $10 \Msunpyr$ in each of the clusters.}
    \label{fig:6cmpt_em}
\end{figure*}

Fig.~\ref{fig:6cmpt_em} shows the normalization of each component as a function of temperature for each of the clusters with the two the spectral models, \textsc{apec} and \textsc{spex}. Also shown on the plot are the expected normalization distributions for a cooling flow of $10\Msunpyr$, using the same metallicities as obtained from the spectral fits. These lines vary because the objects lie at different distances and have different metallicities. At 0.862~keV, the objects have normalizations compatible with $10\Msunpyr$, but at lower temperatures the maximum radiative cooling rate is much lower. Note that the SPEX spectral model gives significantly lower normalizations of the 0.544 and 0.272~keV components than APEC.

The spatial broadening of the lines from the coolest gas imply they come from a regions of approximate radii 0.33, 0.57 and 0.44 arcmin, for Abell~262, Abell~3581 and HCG~62, respectively. 

\begin{figure*}
    \includegraphics[width=\textwidth]{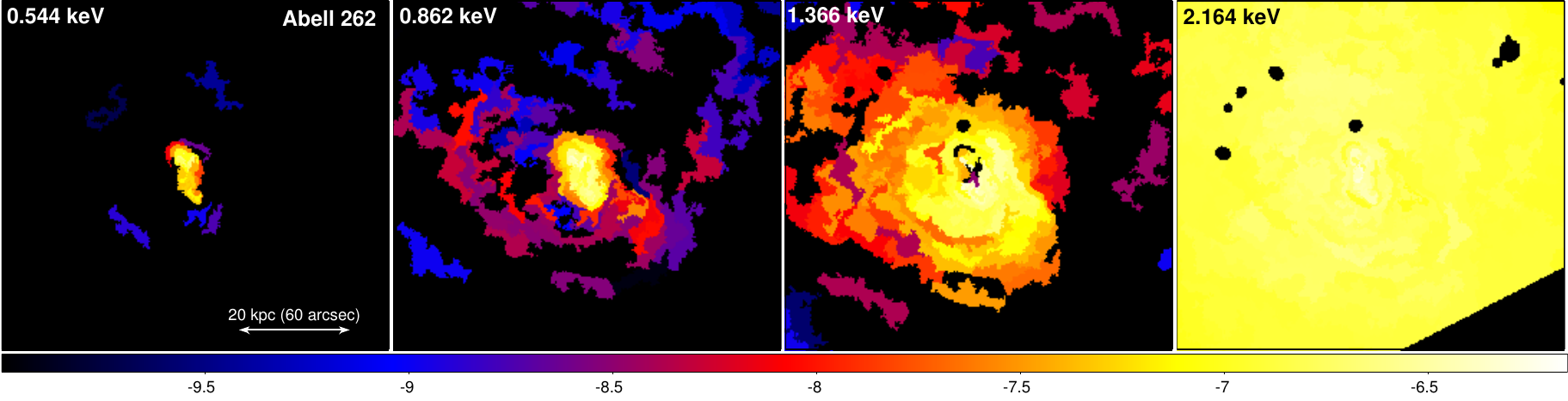}
    \includegraphics[width=\textwidth]{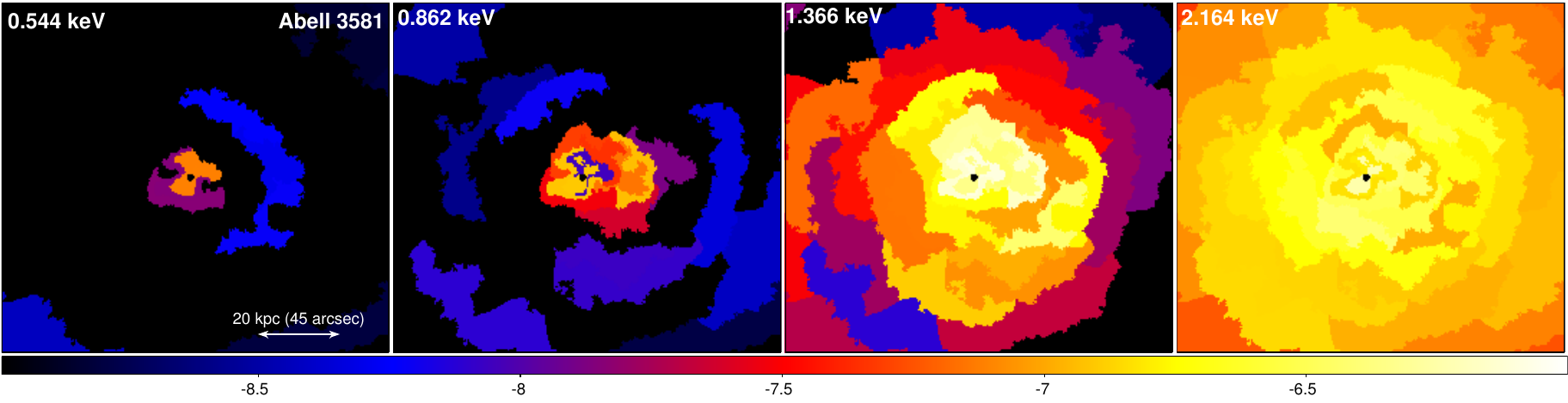}
    \includegraphics[width=\textwidth]{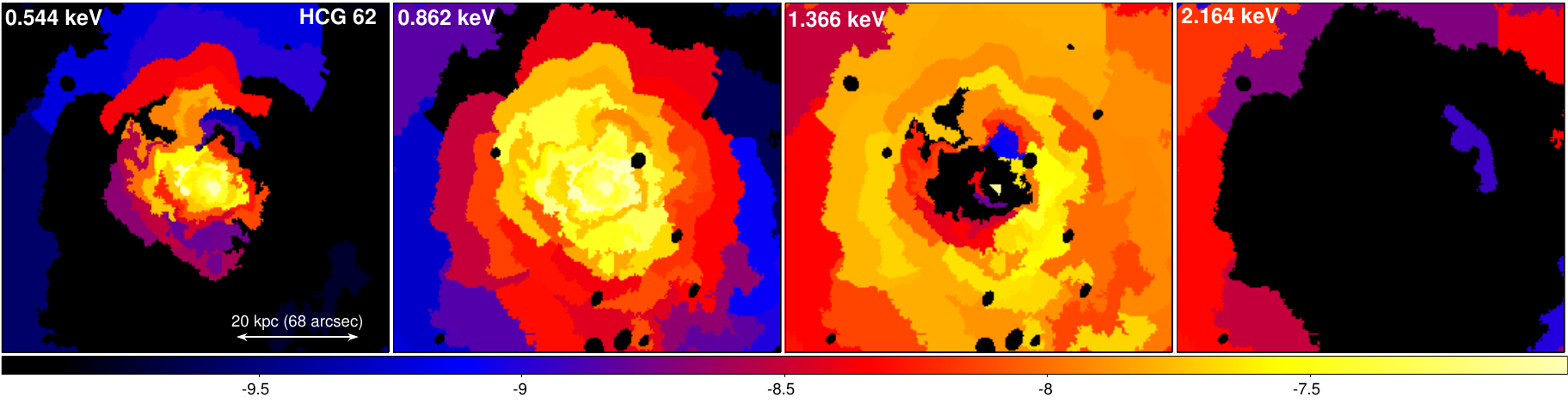}
    \caption{Normalization distributions as a function of temperature from the \emph{Chandra imaging}. The colour scales are log$_{10}$ \textsc{xspec} normalizations per square arcsec. The regions for Abell~262, Abell~3581 and HCG~62 contain a signal to noise ratio of 32, 22 and 22, respectively.}
    \label{fig:emdist}
\end{figure*}

We can also compare these results against the normalization distribution we can measure from \emph{Chandra} data. If we fit a multitemperature model to individual spatial regions we can map the distribution of gas as a function of temperature. We used the Contour Binning algorithm to choose spatial regions from the Chandra data. As the Abell~262 observation had more counts than the other two, we chose a minimum signal to noise ratio of 32 ($\sim 1024$ counts). For the other two objects we used 22 ($\sim 480$ counts). We extracted spectra from each of the regions. We fitted a model to each spectrum made up of 0.544, 0.862, 1.366 and 2.164~keV \textsc{apec} components. We did not include the 0.272 and 1.719 keV components as the data are not sensitive enough to measure these. The metallicity was allowed to vary in each fit, but the metallicities of the temperature components were tied together. The absorbing column density was tied to the Galactic value. The C statistic was minimized in the spectral fitting.

Fig.~\ref{fig:emdist} shows the resulting normalization per unit areas for each of the objects in each of the temperature components. The images show we detect cool temperature components in the core of each cluster as with the RGS data. 
\begin{figure*}
    \includegraphics[width=0.3\textwidth]{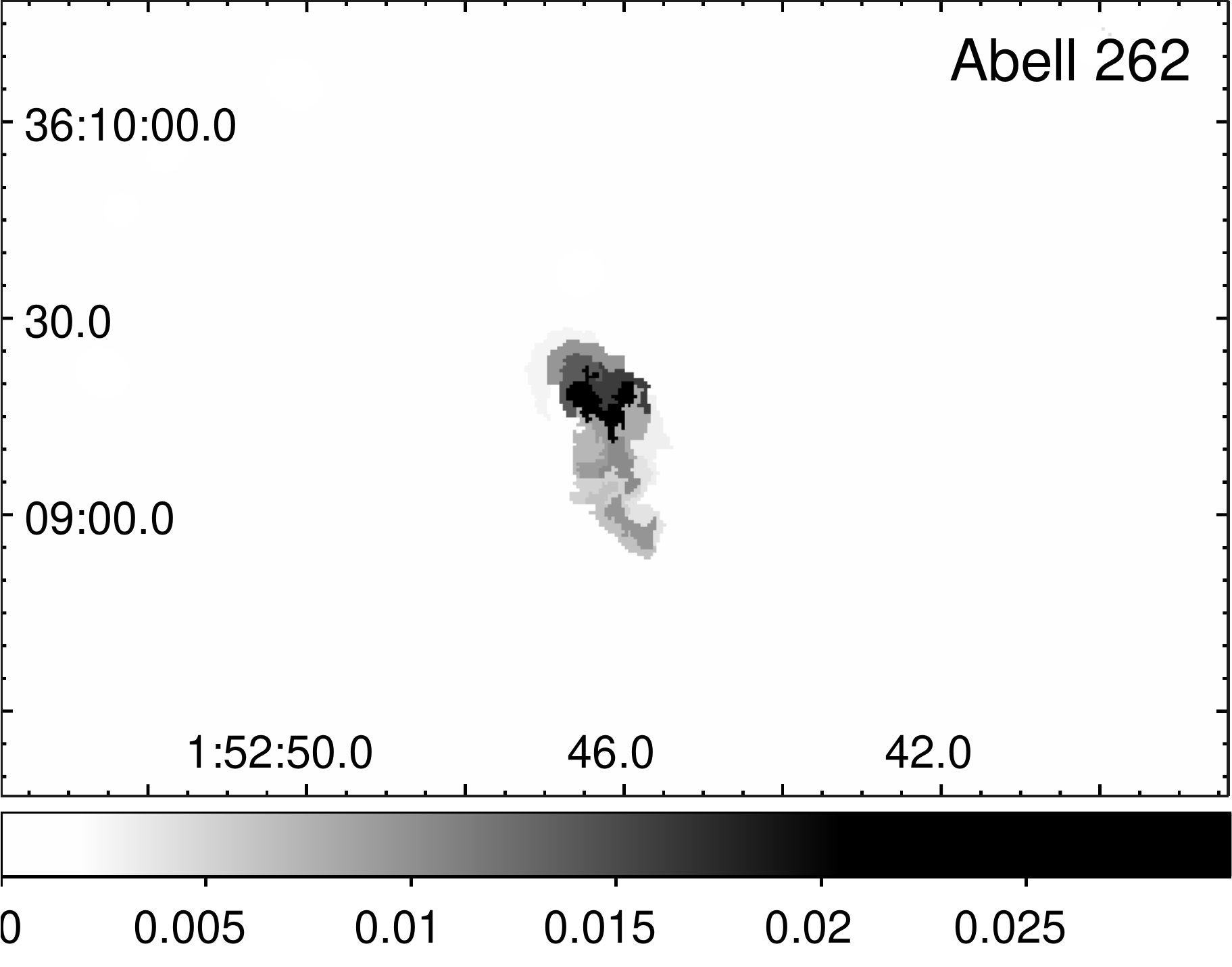}
    \includegraphics[width=0.3\textwidth]{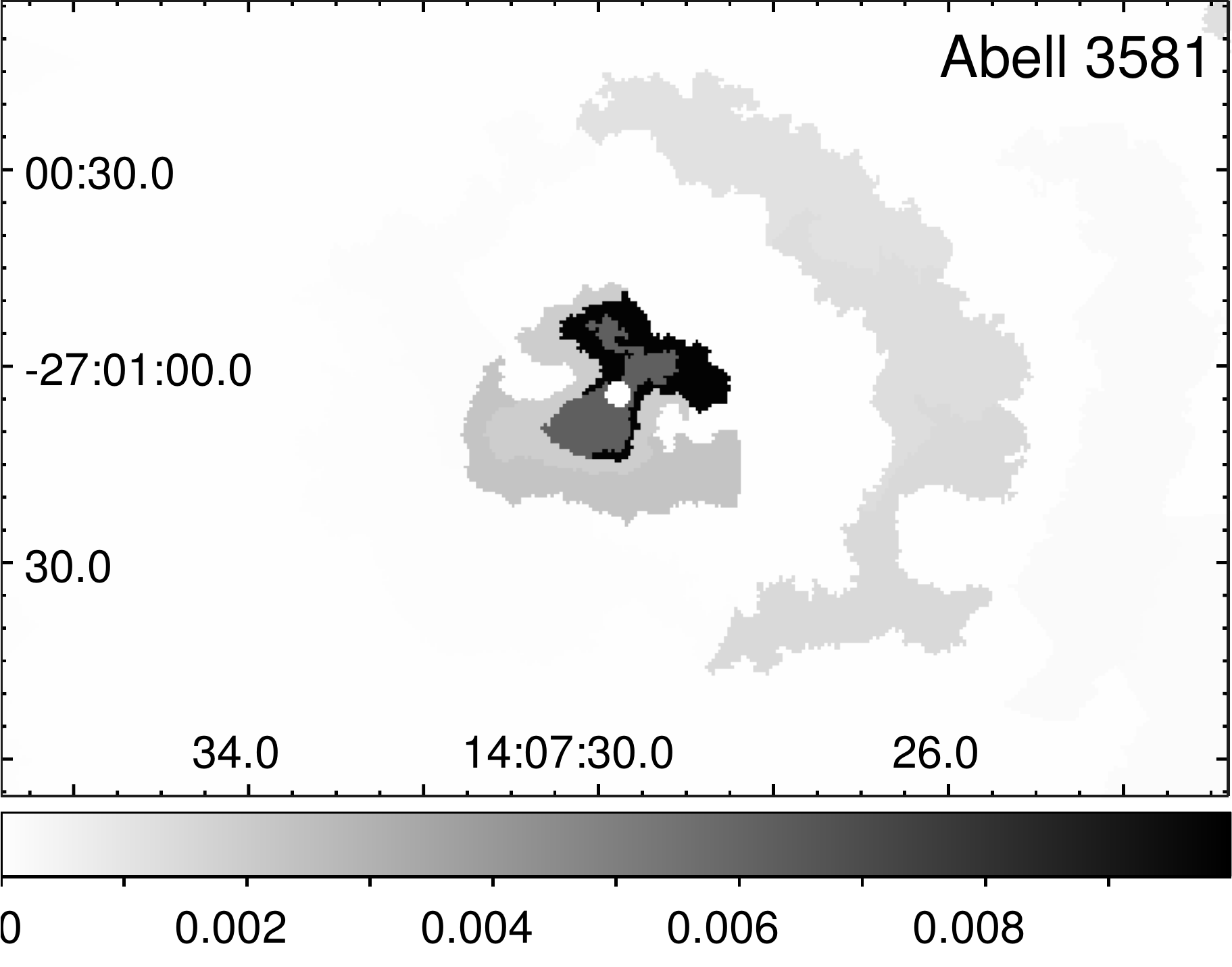}
    \includegraphics[width=0.3\textwidth]{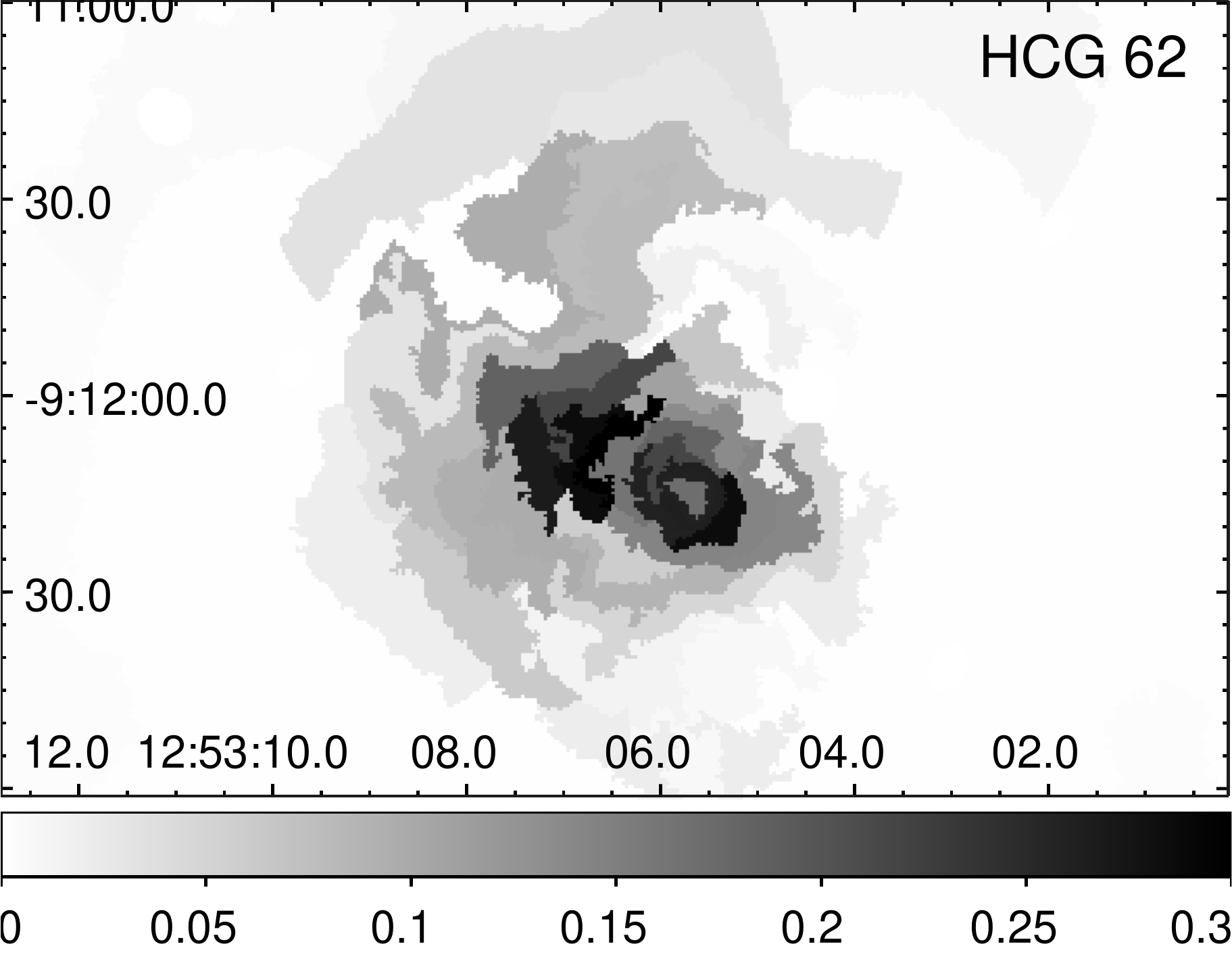}
    \caption{Maps of the volume filling fraction of the 0.544 keV component in the fits to \emph{Chandra} data, calculated from the values in Fig.~\ref{fig:emdist}.}
    \label{fig:vff}
\end{figure*}

The volume filling fraction of the coolest temperature component can be calculated assuming that the components are in pressure equilibrium. The volume filling fraction of component $i$ is
\begin{equation}
    f_i = \frac{ \mathcal{E}_i \: T_i^2 }{\sum_j \mathcal{E}_j \: T_j^2 },
\end{equation}
where $\mathcal{E}_i$ is the normalization of component $i$, $T_i$ is its temperature and $j$ sums over all the components. We show the volume filling fraction for the coolest component in Fig.~\ref{fig:vff}. This fraction is very low for Abell~262 and Abell~3581, indicating this is true multiphase gas. The value in HCG~62 is much larger. The projected temperature in the core of HCG~62 lies between 0.544 and 0.862~keV so the signal in these components there is probably not due to multiphase gas. There could still be some multiphase gas at lower temperatures still.

We also calculate the mass in the 0.544~keV components assuming they are in pressure equilibrium with the pressure of the gas calculated using the \emph{Chandra} deprojection analysis. These values are plotted on Fig~\ref{fig:mass_t}, compared with the distributions from the \emph{Chandra} radial analysis.

\subsubsection{Markov Chain Monte Carlo spectral analysis}
\label{sect:mcmcspectra}
The fixed temperature component models provides a reasonable description of the data. To construct a normalization plot without discrete temperature bins we used a MCMC analysis.

We constructed a model made up of four \textsc{apec} components with the metallicities fixed to be the same. These components were first fixed to be at 0.5, 0.8, 1.2 and 2 keV. The coolest three components shared the same spatial broadening size. The 2~keV component was broadened by a separate value. We fit the model to get initial metallicities, normalizations and spatial broadening sizes.

In the MCMC analysis, we included in the chain the temperatures of the three lowest temperature components, their single broadening size and the metallicities of all the components. To construct the chain we used a custom proposal distribution in \textsc{xspec} which dynamically adjusted a Gaussian proposal distribution for each parameter in every 100 iterations so as to have a repeat fraction close to 75 per~cent during the burn in period. Each parameter was stepped sequentially in the chain in random order. We ran the chain three times for each object. Each chain had a length of $10^6$ and the initial $2\times10^4$ values were discarded. The temperatures were constrained to lie between 0.08 and 4~keV.

\begin{figure}
    \includegraphics[width=\columnwidth]{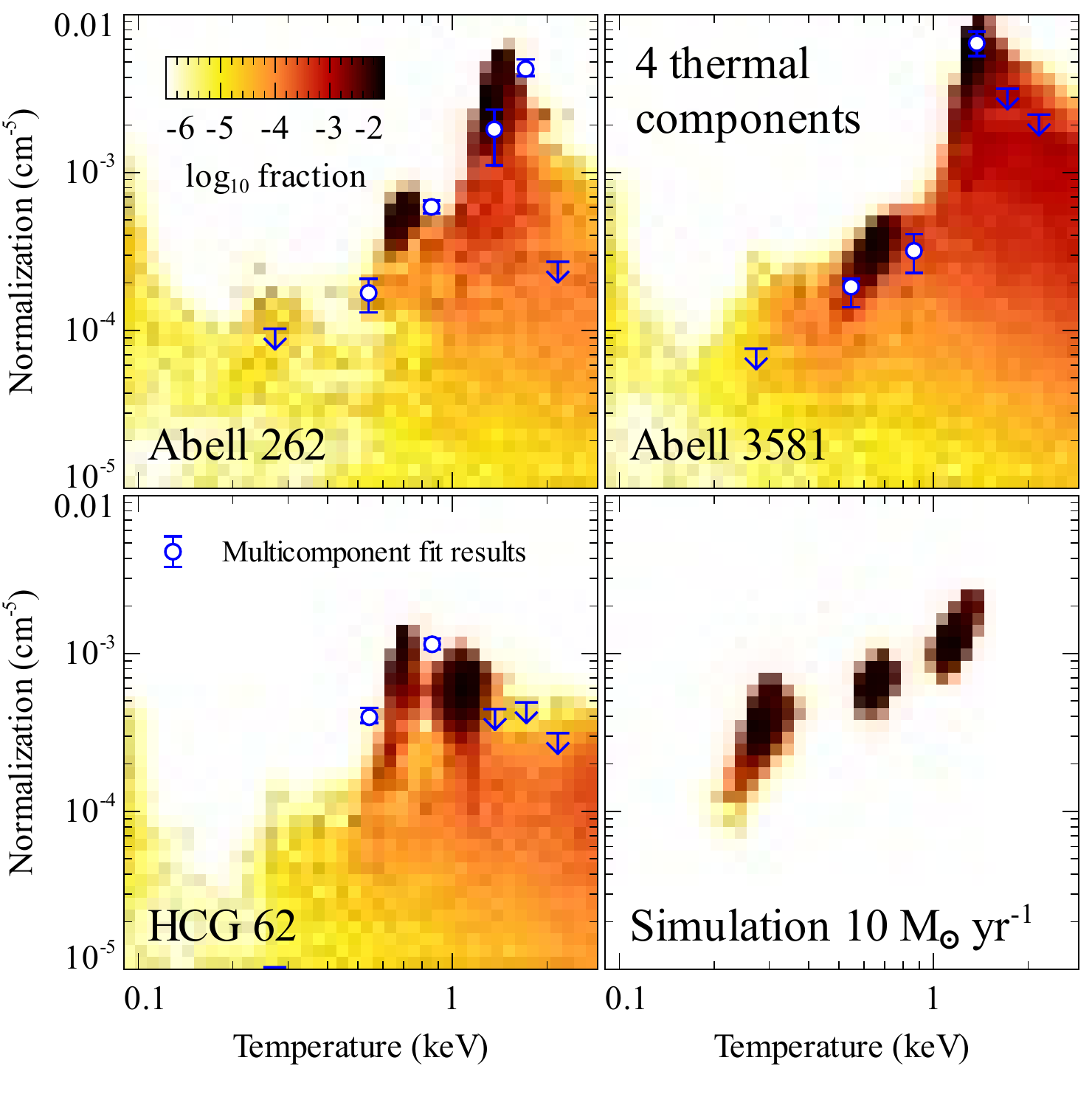}
    \caption{MCMC distributions of temperature and normalization from a multitemperature model with three variable temperature components and one fixed at 2 keV. Also shown is the distribution from a simulated $10\Msunpyr$ cooling flow at the redshift of Abell~262. The points plotted are the results from the 6 component \textsc{vapec} fits in Fig.~\ref{fig:6cmpt_em}}
    \label{fig:mcmcdist}
\end{figure}

Fig.~\ref{fig:mcmcdist} shows the distribution of temperatures and emission measures for the three unfixed temperature components. The fraction shown on the image is the fraction of times one of the free temperature components in the chain was in that particular temperature-normalization bin. There are 40 logarithmic temperature bins between 0.09 and 2.9~keV and 40 logarithmic normalization bins between $10^{-5}$ and $10^{-2}$~cm$^{-5}$ plotted. We also include the results from running the MCMC analysis on a simulated 120~ks dataset for a $10 \Msunpyr$ cooling flow cooling from 2~keV at the redshift of Abell~262 and with $0.6\Zsun$ metallicity. The dataset was simulated with the Abell~262 RGS responses and a single spatial broadening.

These distributions are not completely straightforward to interpret. The results from the simulated dataset show that fitting discrete temperature components to a spectrum generated from continuous distribution yields particular temperatures, rather than a continuous distribution. The gas temperatures which are easily differentiated spectrally in the RGS data are likely to be these particular temperatures.

In the real data, the high likelihood temperature-normalization regions have similar temperatures to the two hottest likely regions from the simulated data. The coolest region in the simulated data (around 0.3~keV) is not seen in the real data. In the real data the third temperature component moves around in the allowed temperature space giving the continuous distribution seen.

The simulated example shows that any ``special'' temperatures that come out of an analysis should be treated with caution. A spectrum generated from a continuous distribution of temperatures will yield particular temperatures under analysis as a matter of course. These particular temperatures will depend on a combination of instrumental response and how spectral features change as a function of temperatures.

The lack of the third component, seen in the cooling flow model simulation results around 0.2 keV, in the real data confirms that we do not observe any gas around these temperatures. We also see very good agreement between the simple 6 component spectral fitting results from Section \ref{sect:multicomp} (Fig.~\ref{fig:6cmpt_em}) which are also plotted as points in Fig.~\ref{fig:mcmcdist}, and these MCMC results.

\begin{figure}
    \includegraphics[width=\columnwidth]{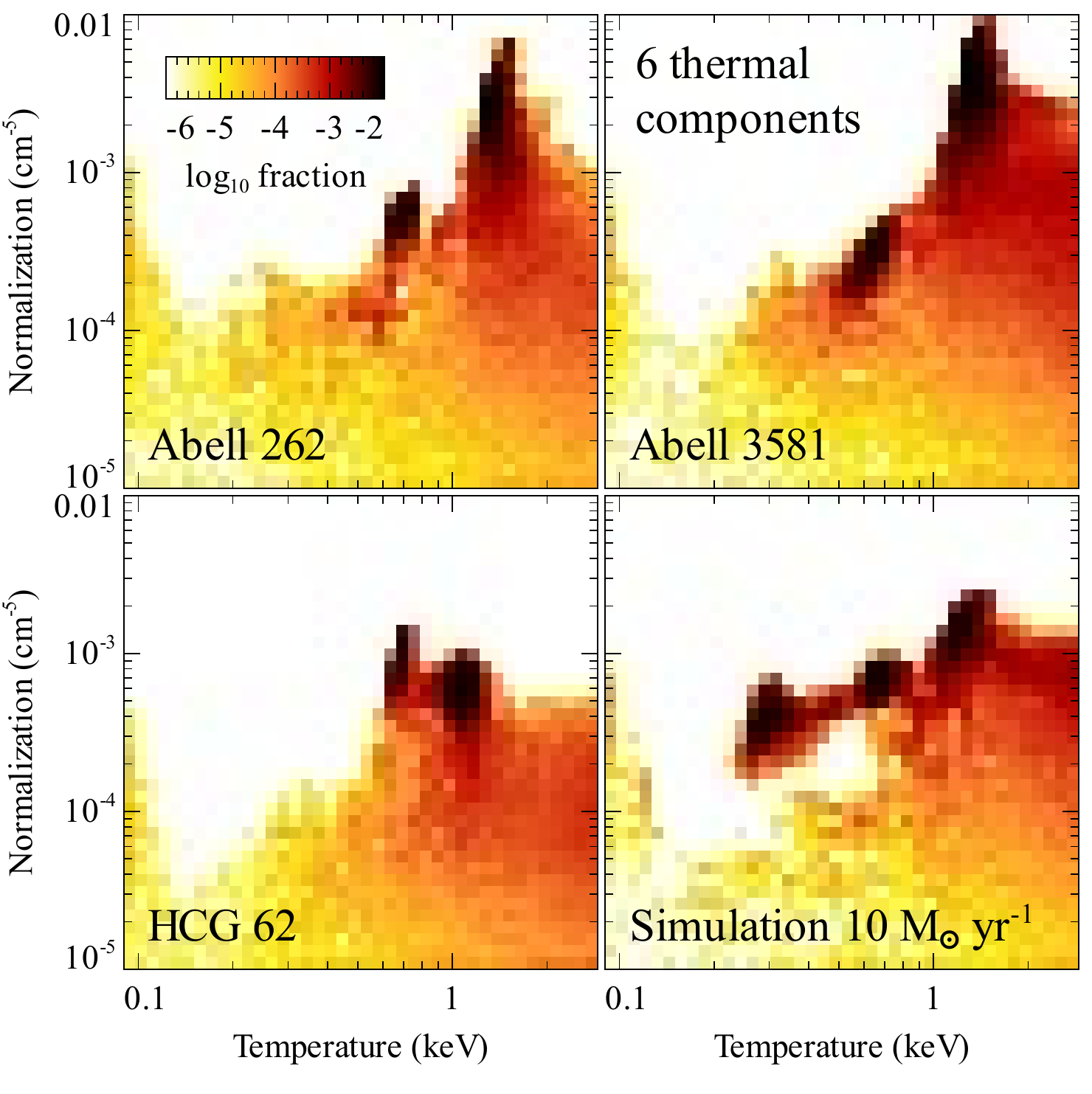}
    \caption{A 6 component MCMC distribution of temperature and normalization. This can be compared against the 4 component results shown in Fig.~\ref{fig:mcmcdist}.}
    \label{fig:mcmcdist6}
\end{figure}

If we increase the number of temperature components from 4 to 6 in the MCMC analysis, we produce plots of temperature and emission measure shown in Fig.~\ref{fig:mcmcdist6}. The results are very similar to the previous plot for real data, showing the results are robust. For the simulated data we see a continuous distribution between the three emission measure peaks.

\subsubsection{Redshift from spectral fitting}
\begin{figure*}
    \includegraphics[width=0.7\textwidth]{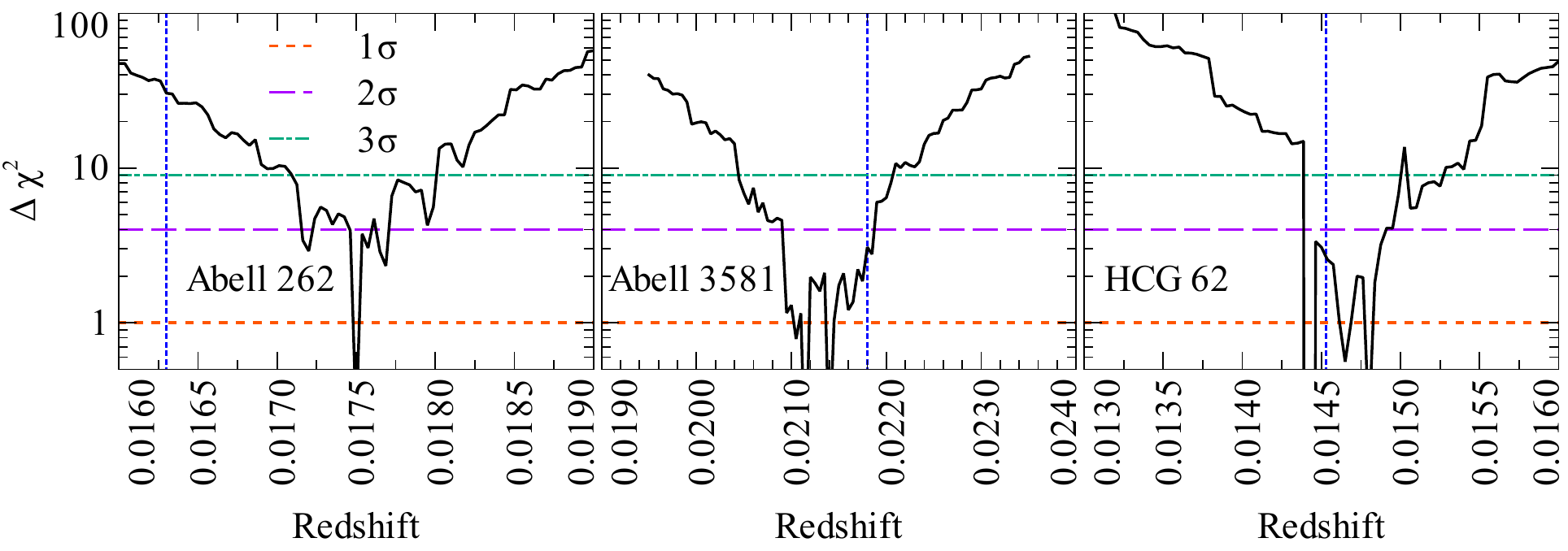}
    \caption{Change in quality of fit as a function of redshift for each of the clusters from the RGS spectra. These results were from the 6 component \textsc{apec} model with free redshift. The vertical line shows the redshifts taken from Table \ref{tab:sample}. The horizontal lines show the $\Delta \chi^2$ values appropriate for confidence regions of 1, 2 and $3\sigma$.}
    \label{fig:redshift}
\end{figure*}

We can also measure the mean redshift of the intracluster medium in each object by spectral fitting. Using the $6\times$\textsc{vapec} model from Section \ref{sect:multicomp}, we allowed the joint redshift of the 6 thermal components to be free.

Fig.~\ref{fig:redshift} shows the change in the quality of the fit ($\chi^2$) as a function of redshift. We find the redshifts we obtain are roughly consistent with the measured optical values in Abell~3581 and HCG~62, but not in Abell 262. There is an offset of around $360 \kmps$ between the two values. Such a shift could be due to in incorrect position for Abell 262 during the processing. However the wavelength shift at 19{\AA} would require a large offset of around 8 arcsec. Therefore it is probably a real offset between the optical measurement and X-ray value. \cite{Sakai94} examine the velocity distribution of the galaxies in Abell~262 as a function of the distance from the cluster centre. Our velocity is within their distribution of velocities, but not at the peak. There is no trend towards our value towards the cluster centre. The D galaxy NGC 708 lies at the peak of the X-ray emission \citep{JonesForman84}, but its velocity does not match our velocity measurement either. There may be a bulk flow of gas in the cluster giving our velocity offset.

We did not use the best fitting redshift of Abell~262 when examining the temperature distribution of the ICM. If we allow the redshift to be free when fitting the six thermal components spectral model (Section \ref{sect:multicomp}), it makes no significant changes to the normalizations of each component. The error bars become slightly larger but there is no systematic shift in the best fitting values.

\subsubsection{Multi-component cooling flow spectral fitting}
\begin{figure}
  \centering
  \includegraphics[width=0.9\columnwidth]{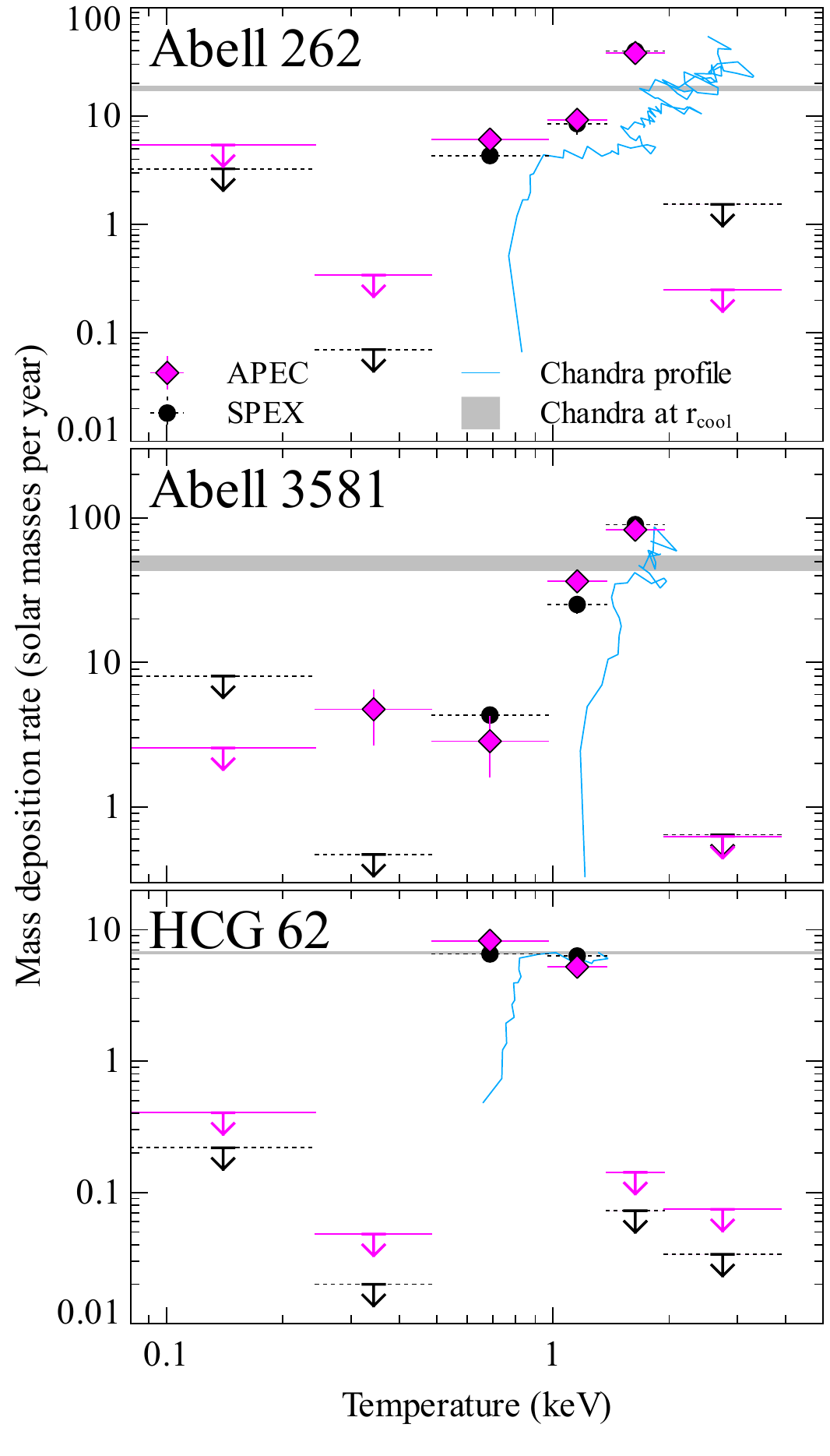}
  \caption{Comparison of mass cooling rates obtained with \textsc{apec} and \textsc{spex} cooling flow models using 6 temperature bins. The shaded regions show mass deposition rates derived from \emph{Chandra} surface brightness deprojection (Table \ref{tab:deproj}). The solid line was created by plotting temperatures against cumulative mass deposition rates from Fig.~\ref{fig:chandraprofiles}, with the temperatures smoothed with a 3 bin sliding average.}
  \label{fig:vmcflow_apec_spex}
\end{figure}

\begin{figure}
  \centering
  \includegraphics[width=0.9\columnwidth]{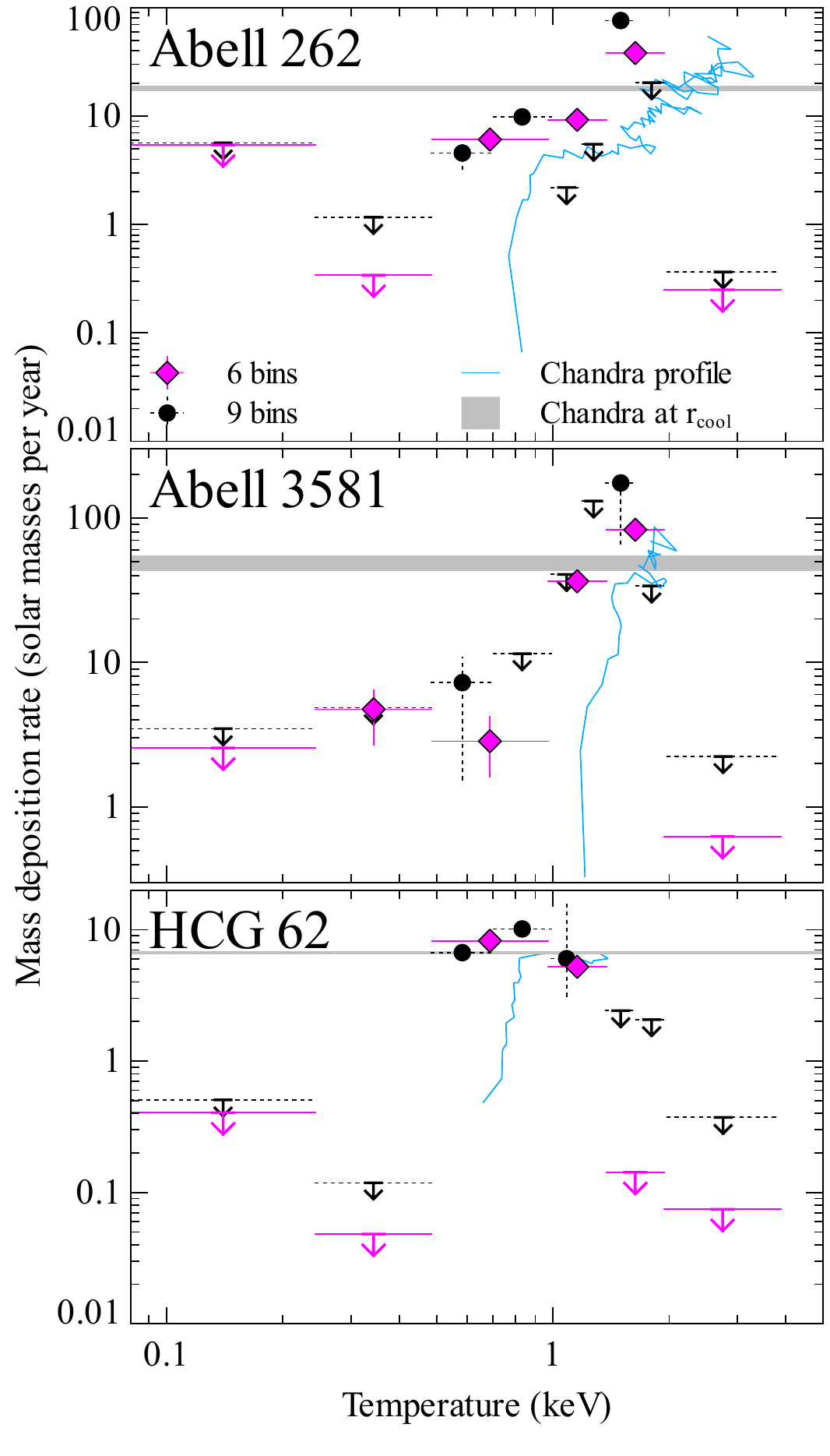}
  \caption{Comparison of mass cooling rates obtained with \textsc{apec} cooling flow models using 6 or 9 temperature bins. The shaded regions show mass deposition rates derived from \emph{Chandra} surface brightness deprojection (Table \ref{tab:deproj}). The solid line shows temperatures plotted against cumulative mass deposition rates from Fig.~\ref{fig:chandraprofiles}.}
  \label{fig:vmcflow_6_9}
\end{figure}

As an alternative to the multitemperature model from Section \ref{sect:multicomp}, we can fit models which parametrize the amount of gas which could be cooling as a function of temperature. We used a model made up of several components which modelled gas cooling through different temperature ranges. We firstly used six bins between temperatures of 3.88 to 1.94, 1.94 to 1.37, 1.37 to 0.97, 0.97 to 0.49, 0.49 to 0.24 and 0.24 to 0.080 keV.

Fig.~\ref{fig:vmcflow_apec_spex} compares the mass cooling rates obtained using the \textsc{apec} spectral model and the \textsc{spex} spectral model. The results are fairly similar between the two models except in the 0.49 to 0.24 bin and for the different limits in the coolest temperature bin. Also shown on the plot are the cumulative mass deposition rates as a function of temperature obtained with the \emph{Chandra} surface brightness deprojection (Fig.~\ref{fig:chandraprofiles}). The cumulative mass deposition rate at the cooling radius is shown as a shaded bar (the values are taken from Table \ref{tab:deproj}). There is quite a good agreement between the shape of the RGS temperature-mass cooling rate relation and the \emph{Chandra} profile version, except for an offset in temperature between them.

Fig.~\ref{fig:vmcflow_6_9} shows a comparison between the results using the \textsc{apec} model with 6 bins and one with 9 temperature bins. The results are consistent between the two different temperature binnings. As expected, the constraints are relaxed with more temperature bins.

\begin{figure}
    \centering
    \includegraphics[width=\columnwidth]{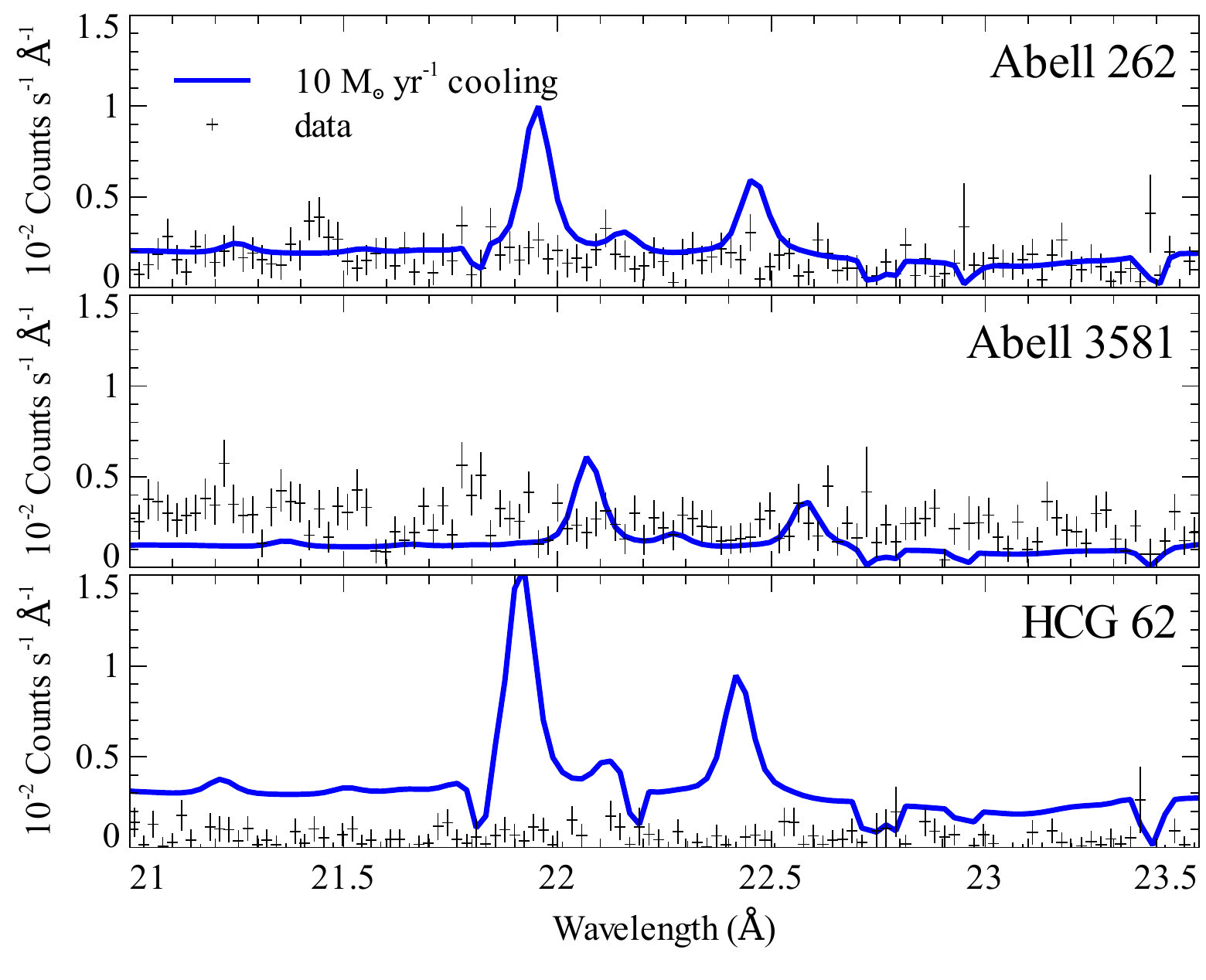}
    \caption{RGS spectra around the expected position of the O\textsc{vii} lines. These spectra have not been fluxed but have been background subtracted. The thick solid line is a model cooling flow spectrum cooling from 3~keV to zero temperature at the rate of $10 \Msunpyr$ (at Solar metallicity).}
    \label{fig:oviilines}
\end{figure}

In Fig.~\ref{fig:oviilines} we show unfluxed background subtracted spectra between 21 and 23.5{\AA}, showing the region where O~\textsc{vii} lines should be observed if they exist. Also shown are model spectra for a $10 \Msunpyr$ cooling flow for Solar metallicity material. O~\textsc{vii} is one of the strongest indicators of gas around 0.25~keV in our wavelength range, however it is not one of the main coolants. Therefore the strength of the line will depend on the metallicity of oxygen and amount of cool gas.

\begin{figure}
    \centering
    \includegraphics[width=0.8\columnwidth]{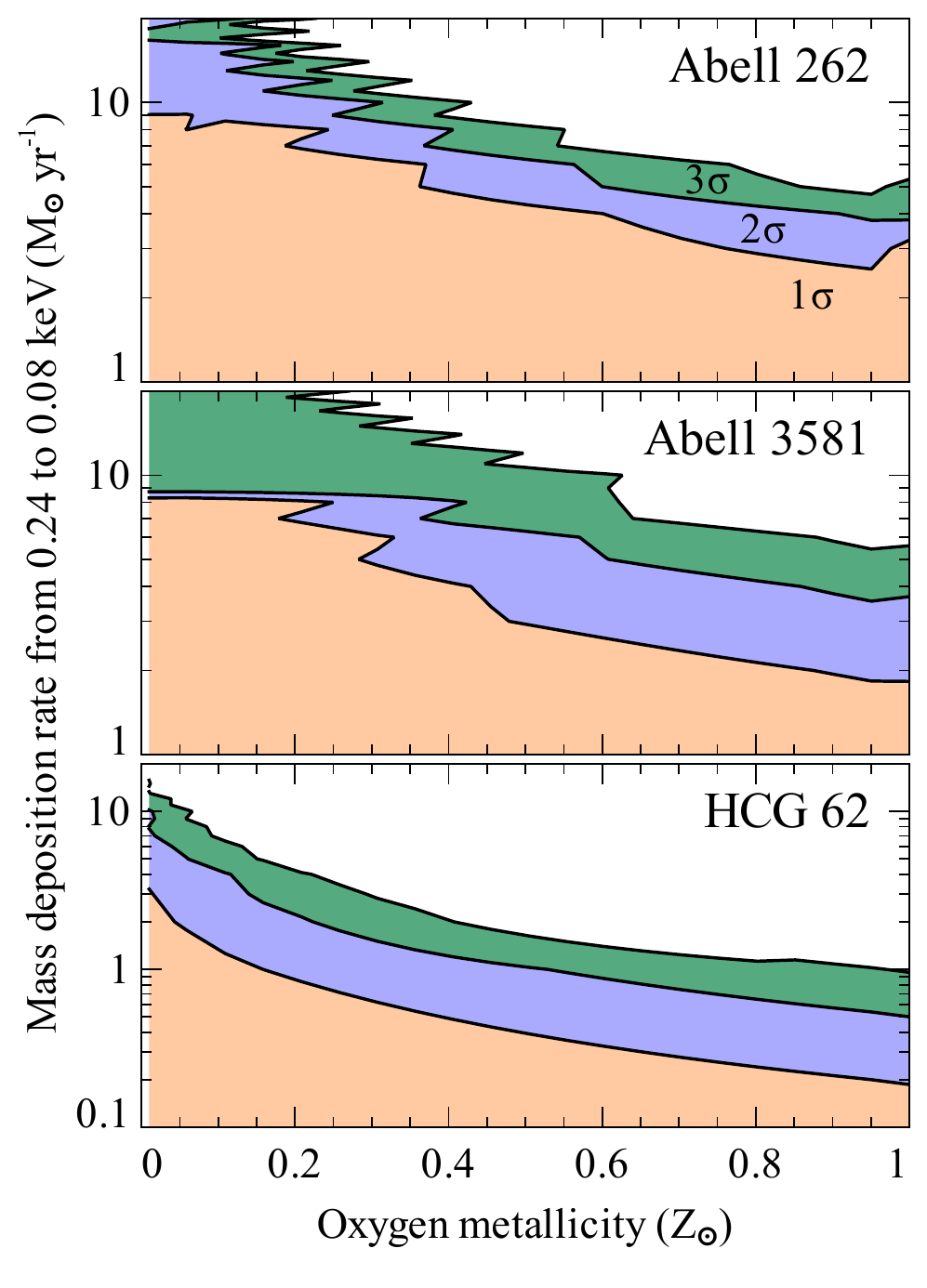}
    \caption{Confidence contours on limits mass deposition rate in the 0.24 to 0.08~keV temperature range as a function of oxygen metallicity. This was measured with the \textsc{apec} model (see Fig.~\ref{fig:vmcflow_apec_spex}).}
    \label{fig:omdot}
\end{figure}

\begin{figure*}
    \includegraphics[width=0.75\textwidth]{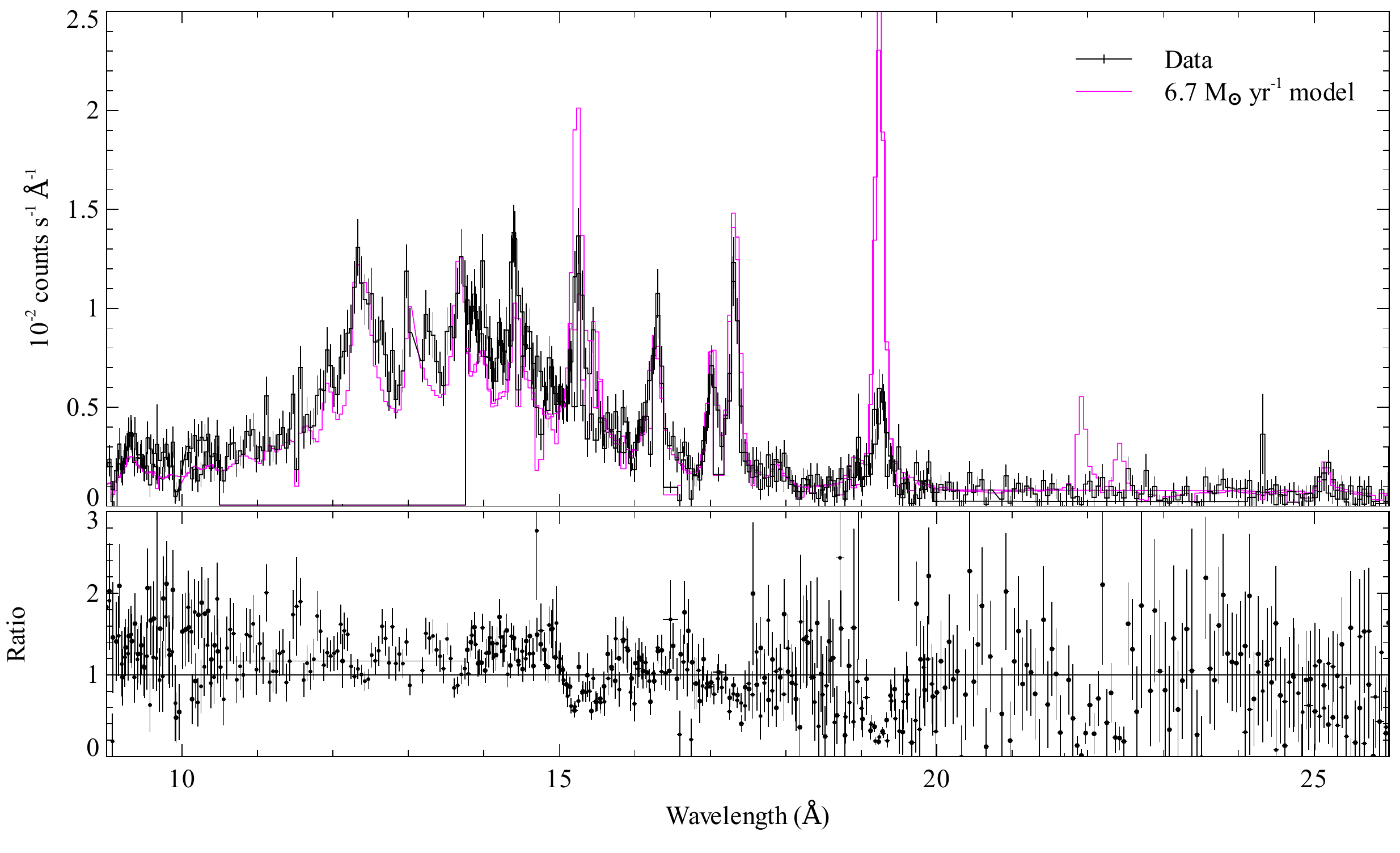}
    \caption{HCG 62 spectrum and comparison with a $6.7\Msunpyr$ cooling flow (cooling from 1.2 to 0.0808~keV with $0.5\Zsun$ metallicity).}
    \label{fig:hcg62_coolspec}
\end{figure*}

We can examine how the constraint on the upper limit on the amount of gas cooling in the lower temperature bin is affected by the oxygen metallicity.  Lower oxygen abundances could potentially be caused by a relative deficit of Type II supernovae in the core of the clusters. In Fig.~\ref{fig:omdot} we show confidence contours for the mass cooling rate as a function of oxygen metallicity, for the lowest temperature bin in the \textsc{apec} cooling model (see the best fitting values in Fig.~\ref{fig:vmcflow_apec_spex}). The plots show that we could have larger amounts of cooling at low temperatures if oxygen metallicities are depleted at these temperatures, but very low metallicities are required to match the cooling rates seen at higher temperatures.

Fig.~\ref{fig:hcg62_coolspec} shows a comparison of the HCG~62 data with a model cooling to the minimum temperature at $6.7\Msunpyr$ at $0.5\Zsun$. This model is broadly consistent with the Fe~\textsc{xvii} emission lines close to 15 and 17{\AA}, but is inconsistent with the O~\textsc{viii} emission line at 19{\AA} and the O~\textsc{vii} lines at 22.6 and 22.1{\AA}. Reducing the oxygen abundance would help the fit, but we would need unrealistically low values ($\sim 0.05\Zsun$) to match the data.

\subsection{Line ratios}

\begin{table*}
    \caption{Measured line fluxes ($10^{-14}$ \ergcmsqps). These are measured from a 95 per~cent extraction region. Uncertainties and limits are $1\sigma$.}
    \begin{tabular}{lcccccc}
\hline
Cluster      & \multicolumn{4}{|c|}{Line} \\
\hline

             & \multicolumn{4}{|c|}{Fe \textsc{xvii}}
             & \multicolumn{2}{|c|}{O \textsc{vii}}\\
$\lambda$ (\AA) & 15.01 (3C)          & 15.26 (3D)             & 16.78 (3F)       & 17.05 (3G+M2)
             & 21.60                  & 22.10                  \\
\hline
Abell 262    & $2.62 \pm 0.41$        & $1.15 \pm 0.32$        & $1.07 \pm 0.35$ & $3.43 \pm 0.34$
             & $<0.42$                & $<0.25$                \\
Abell~3581   & $1.99^{+0.32}_{-0.49}$ & $0.77^{+0.26}_{-0.38}$ & $0.51 \pm 0.35$ & $2.11 \pm 0.33$
             & $<0.30$                & $<0.35$                \\
HCG~62       & $4.74 \pm 0.40$        & $1.56 \pm 0.25$        & $3.23 \pm 0.32$ & $5.33 \pm 0.30$
             & $<0.13$                & $<0.17$                \\
\hline
    \end{tabular}
    \label{tab:linestrengths}
\end{table*}

\begin{figure}
    \centering
    \includegraphics[width=0.8\columnwidth]{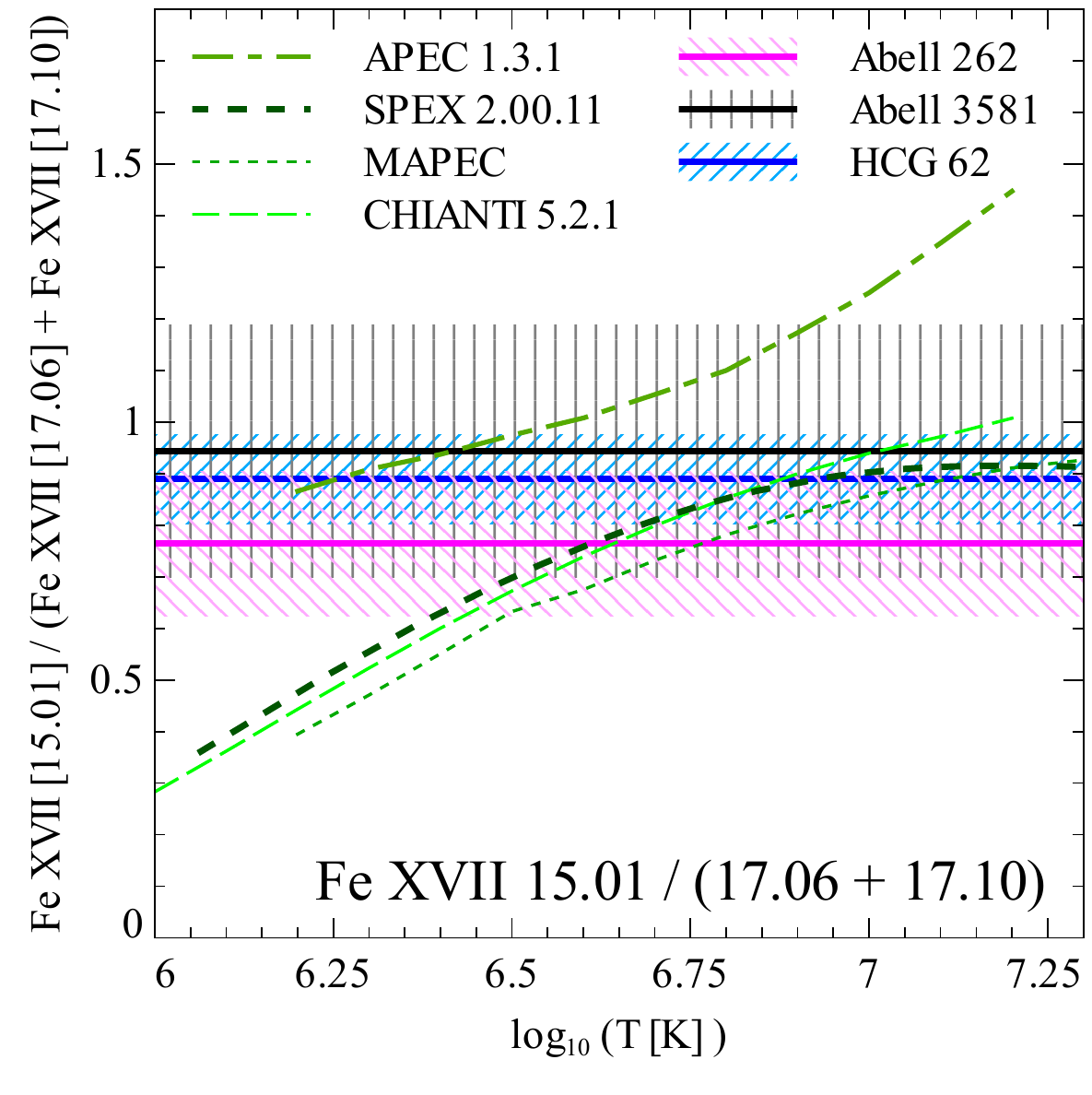}
    \caption{Measured Fe\textsc{xvii} line ratios (in terms of energy flux) compared to theoretical models. The ratio measured in a cluster is shown by a line surrounded by a shaded region which represents the $1\sigma$ uncertainty.}
    \label{fig:lineratio}
\end{figure}

A different way of examining the temperature distribution is to look at temperature sensitive emission lines, such as the different Fe \textsc{xvii} emission lines. In our spectra the strongest lines are the 3C line at {15.01\AA} (a $2p_{1/2}-3d_{3/2}$ transition) and the combined 3G and M2 lines at {17.1\AA} (both $2p-3s$ transitions). The 3F line at {16.8\AA} (another $2p-3s$ transition) is also clearly seen in HCG~62. The 3D line at {15.26\AA} is blurred into the 3C lines. The useful 3C/3D ratio cannot be used because of the blurring and low signal to noise of the 3D line. The 3C line can also be subject to resonant scattering due to its oscillator strength.

We can measure the 3C/(3G+M2) ratio. Unfortunately there are disagreements between theory and experiment over how this ratio varies as a function of temperature \citep{Beiersdorfer02,Beiersdorfer04}. In addition, the cluster is likely to contain a range of gas temperature, so the line ratio will not come from a single temperature component. Nevertheless, we measured the strength of the emission lines.

These results were measured by first taking the 6 component \textsc{apec} spectral fit from Section \ref{sect:specfitting}. We then modified the \textsc{apec} table of lines at each temperature to make the emissivity of the Fe~\textsc{xvii} and O~\textsc{vii} lines to be zero. We then added four redshifted Gaussian components for the Fe~\textsc{xvii} lines and two for the O~\textsc{vii} lines. The width of the Fe~\textsc{xvii} lines were constrained to be the same. The width of the O~\textsc{vii} lines were set to be zero. We fit the spectrum over the same range of wavelength as in Section \ref{sect:specfitting}. The line strengths or upper limits we obtained are shown in Table \ref{tab:linestrengths}. This spectral fitting procedure was used to make sure that we had a good model for the continuum around the lines. A simple powerlaw model was not sufficient, giving results dependent on the range of wavelength fitted.

The 3C/(3G+M2) ratio for each cluster is plotted in Fig.~\ref{fig:lineratio}. In addition we plot the line ratio as a function of temperature from \textsc{apec}, \textsc{spex}, \textsc{mapec} \citep{Gu07} and \textsc{chianti} \citep{Dere97,Landi06}. Most of the models show similar ratios as a function of temperature, except for \textsc{apec}.

\cite{DoronBehar02} presented a detailed study of the Fe \textsc{xvii} emission line ratios. They included the effects of radiative recombination, dielectronic recombination, resonant excitation and inner-shell collisional ionization, in addition to the collisional excitation. They found that the 2p-3s lines at 16.78, 17.05 and {17.10\AA} were enhanced by 25, 30 and 55 per cent, respectively. \textsc{apec} is the outlying spectral model in Fig.~\ref{fig:lineratio} and does not include the results of \cite{DoronBehar02}, though work is being done to improve the Fe~\textsc{xvii} line strengths (R. Smith, private communication). \textsc{spex} does include the results of \cite{DoronBehar02}. The line ratios are not strong functions of temperature above $\sim 0.5$~keV, so it is difficult to get an exact temperature measurement using Fe\textsc{xvii}.

The \textsc{apec} model, generally produces a better quality fit overall compared to \textsc{spex}. The two component thermal fit ($2\times$\textsc{vapec}) for Abell~262 has a $\chi^2$ lower by 100 compared to the \textsc{spex} model. We therefore use \textsc{apec} for most of the analysis in this paper.

We do not see any evidence for resonant scattering. However, we would not expect to see this necessarily, as the region where any scattering would take place is likely to be inside our spectral extraction region. The total spectrum from this region would remain the same if there is only scattering within it. Analysis of the line profiles is required to look for scattering (e.g. \citealt{Xu02}).

\section{Smooth particle inference modelling}
\label{sect:smoothparticle}
Rather than the traditional analysis we have performed in Section~\ref{sect:rgsanalysis}, we have also applied a smooth particle inference analysis method to the \emph{XMM-Newton} data, as described in \cite{Peterson07}.

This smooth particle analysis is more complete than the MCMC procedure we outlined in Section \ref{sect:mcmcspectra}. In that analysis we attempted to model the RGS spectrum with just the sum of 4 or 6 spectral components. Here the whole intracluster medium is modelled as a set of X-ray emitting smoothed particles. Each particle is described by a set of parameters including temperature, metal abundance, size and location. Also included in the model are components which account for the instrumental and cosmic X-ray backgrounds. The X-ray photons from the model are then propagated through the response of the X-ray detector and compared to the real data using a MCMC. Once the chain has converged properties about the cluster can be extracted from the chain.

This method is powerful as it can be applied to data from different kinds of instruments, such as gratings and CCD imaging spectroscopy. It builds up a model which fits the spatial and spectral data simultaneously.

We use particles which are modelled as Gaussians spatially, with parameters for the temperature, metal abundance, redshift, normalization, absorbing column density and position and width on the sky. The abundances are restricted to lie between 0 and $2\Zsun$. The Gaussian widths of the particles can lie between 0.03 and 4.08~arcmin. The prior for the distribution of particle temperatures is constant in $-1 \le \mathrm{log}_{10} (kT/\mathrm{keV}) \le 1$. The redshift of the particles and their absorbing column density are fixed. An isothermal absorbed \textsc{mekal} model is used to generate the X-ray spectrum from each particle.

The soft X-ray background is modelled as a spatially uniform unabsorbed \textsc{mekal} component fixed at 0.15~keV (for Abell~262 and Abell~3581) and 0.3 keV (for HCG~62). The component is fixed at $z=0$ and Solar metallicity. The harder X-ray background, produced primarily by unresolved extragalactic sources, is modelled as a $\Gamma=1.47$ spatially-uniform powerlaw, with Galactic absorption. The detector background is also modelled by emission lines representing the contribution from high energy particles and electronic noise, modelled as an exponential. The detector noise is fixed between the different observations.

The photons emitted by the particle and background models are propagated through the RGS and EPIC responses and compared with the input data. The fraction of photons emitted by each component and the relative normalizations of the models are free parameters. Energy cuts of 0.3 to 10 (EPIC MOS), 1.1 to 10 (EPIC PN) and 0.2 to 2.5~keV (RGS) are applied to the photons. The MCMC is run for 1\,000 (EPIC) or 2\,500 (RGS) iterations. The output of every iteration in the chain after 200 is used to produce the set of model parameters. For this analysis we examine only the longer \emph{XMM-Newton} datasets for each object (observations 0504780101, 050478301 and 0504780501), except for the EPIC observation of Abell~262, where we use 0504780201 due to the flaring in the longer observation 0504780101.

\begin{figure}
  \centering
  \includegraphics[width=\columnwidth]{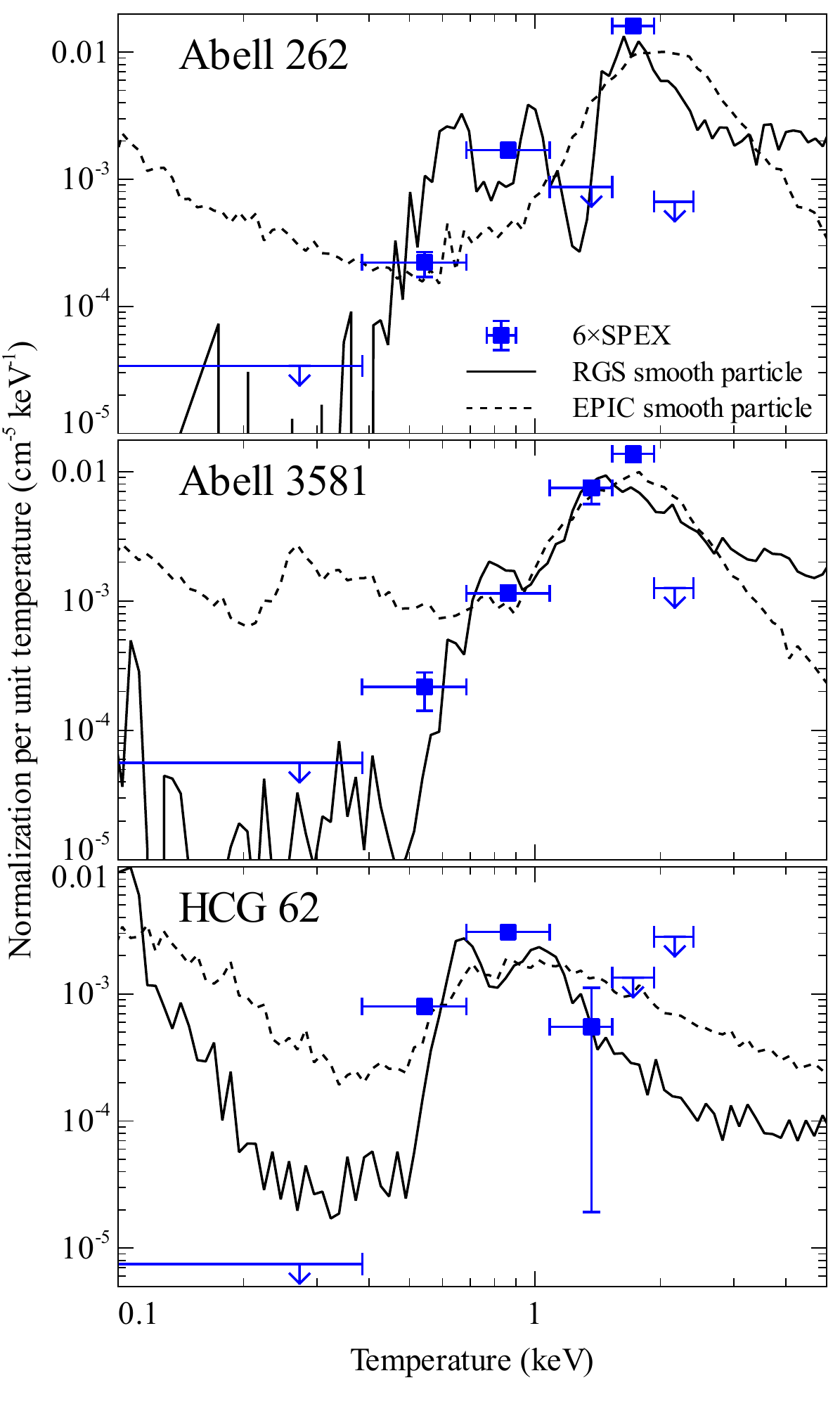}
  \caption{Results from a smooth particle modelling of the clusters, shown as solid (RGS) and dotted (EPIC) lines. Also plotted for comparison are the simple 6 component \textsc{spex} spectral fitting results from Fig.~\ref{fig:6cmpt_em} (plotted as points). Emission measures have been divided by temperature bin widths.}
  \label{fig:smoothparticle}
\end{figure}

In Fig.~\ref{fig:smoothparticle} are shown the resultant distribution of particle temperatures from the MCMC chains. The total normalization of the particles in each logarithmic temperature bin is plotted divided by the bin width. Also plotted for comparison of the 6 component \textsc{spex} results from  Fig.~\ref{fig:6cmpt_em}, divided by an approximate bin width (there are no exact bin boundaries between the temperature components or at the edges of the distribution). We plot the \textsc{spex} spectral fitting results as the spectral model is very similar to the \textsc{mekal} model used in the smoothed particle inference analysis.

Below 2 keV temperature, there is very agreement between the RGS smooth particle results and the simple \textsc{spex} spectral fitting. At high temperatures the results disagree as the simple spectral fitting only includes components up to 2.164~keV in temperature and examines the RGS data from only the central region of the cluster. The EPIC instruments are more sensitive to hot gas than the RGS detectors and so it is not surprising these results disagree with the RGS smooth particle values at high temperatures. Below 0.5~keV there is little sensitivity to cold gas in the EPIC detectors.

\section{Discussion}
\subsection{The nature of the coolest X-ray emitting gas}
\begin{figure}
    \centering
    \includegraphics[width=\columnwidth]{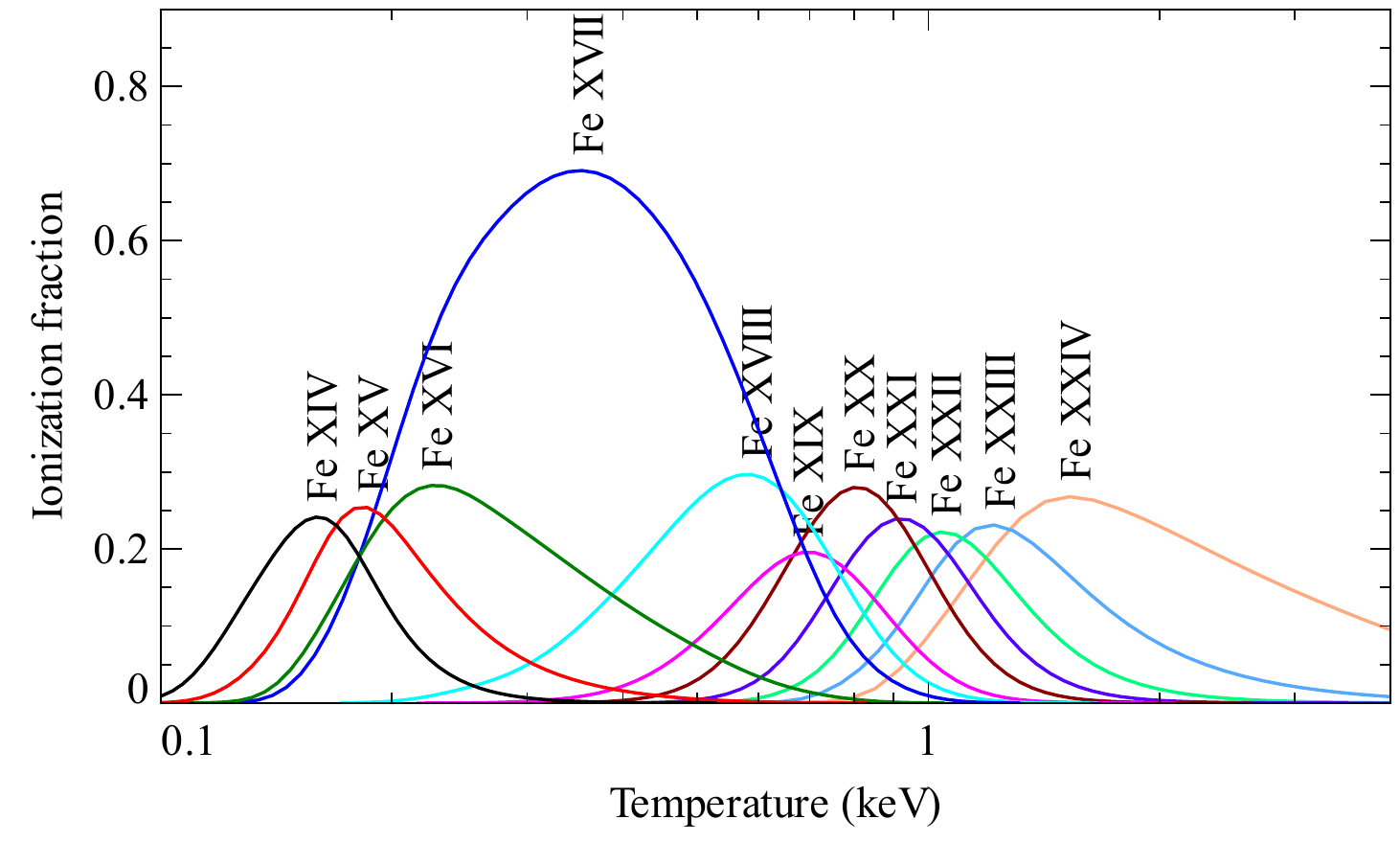}
    \caption{Ionization fraction of different iron ions as a function of temperature. Data are taken from \protect\cite{Mazzotta98}.}
    \label{fig:ionizationfe}
\end{figure}

Purely from the existence of Fe~\textsc{xvii} emission lines in our spectra of these objects we can see that they contain X-ray emitting gas somewhere between approximately 0.15 and 0.80~keV (Fig.~\ref{fig:ionizationfe}). Using spectral fitting, the \textsc{apec} and \textsc{spex} models indicate that there is significant gas at around 0.544~keV (Fig.~\ref{fig:6cmpt_em}).

We can calculate the implied cooling time of the coolest component using its measured emission measure. We assume that the cool gas is in pressure equilibrium with its surroundings, using the peak thermal pressure from the \emph{Chandra} profiles (Fig.~\ref{fig:chandraprofiles}). Taking this pressure and temperature we compute the density and with the fitted metallicities we estimate the cooling time. The cooling times are shown in Fig.~\ref{fig:chandraprofiles}. From the density and the emission measure we can also estimate the volume of the coolest 0.544~keV component. These volumes are around 74, 380 and 612 kpc$^3$ for Abell~262, Abell~3581 and HCG~62, respectively. Using the width of the spectral lines we can also estimate the area on the sky from which they are emitted (Section \ref{sect:specfitting}). If the gas is volume filling, the volume should be compatible with the area. However, from the line widths, the implied volume filling fractions for the 0.544~keV component of 6, 3 and 31 per cent, for Abell~262, Abell~3581 and HCG~62, respectively.

Single phase analyses of CCD data miss this very coldest gas. An intermediate temperature between the volume filling component and the cold blobs is obtained. This can be seen in the discrepancy between the \emph{Chandra} profiles and the RGS values in Figures \ref{fig:vmcflow_apec_spex} and \ref{fig:vmcflow_6_9}. In the Perseus cluster multiphase analysis is required to detect the filaments spectrally (figure 12 in \citealt{FabianPer06}) and in 2A~0335+096, where the multiphase analysis was required to reveal the morphology of the coldest X-ray emitting gas \citep{Sanders2A033509}.

The \emph{Chandra} volume filling fraction maps (Fig.~\ref{fig:vff}) of the coolest components give very similar results to the calculations from the RGS results. The small volume filling factor and spatial extent (from \emph{Chandra} maps and from the line widths in the RGS data) for Abell~262 and Abell~3581 (Fig.~\ref{fig:emdist}) suggests that these components consist of small cool blobs in a hotter medium. It is a true multiphase gas. In HCG~62 the volume filling factor of the 0.544~keV component is much larger (around 30 per cent). This temperature is also close to the mean projected temperature from the cluster, so we are probably detecting the volume filling phase (which lies in temperature between 0.544 and 0.862~keV). The cool blobs in Abell~262 and Abell~3581 are likely to be similar to those found in other clusters, such as 2A~0335+096.

The temperature distribution in HCG~62 is also narrower than in the other two objects, as seen from spectral fitting (Figures \ref{fig:6cmpt_em} and \ref{fig:vmcflow_apec_spex}), the simple MCMC analysis (Fig.~\ref{fig:mcmcdist}) and the full smooth particle inference analysis (Fig. \ref{fig:smoothparticle}).

\subsection{The distribution of gas temperature}
\begin{figure}
 \centering
 \includegraphics[width=0.9\columnwidth]{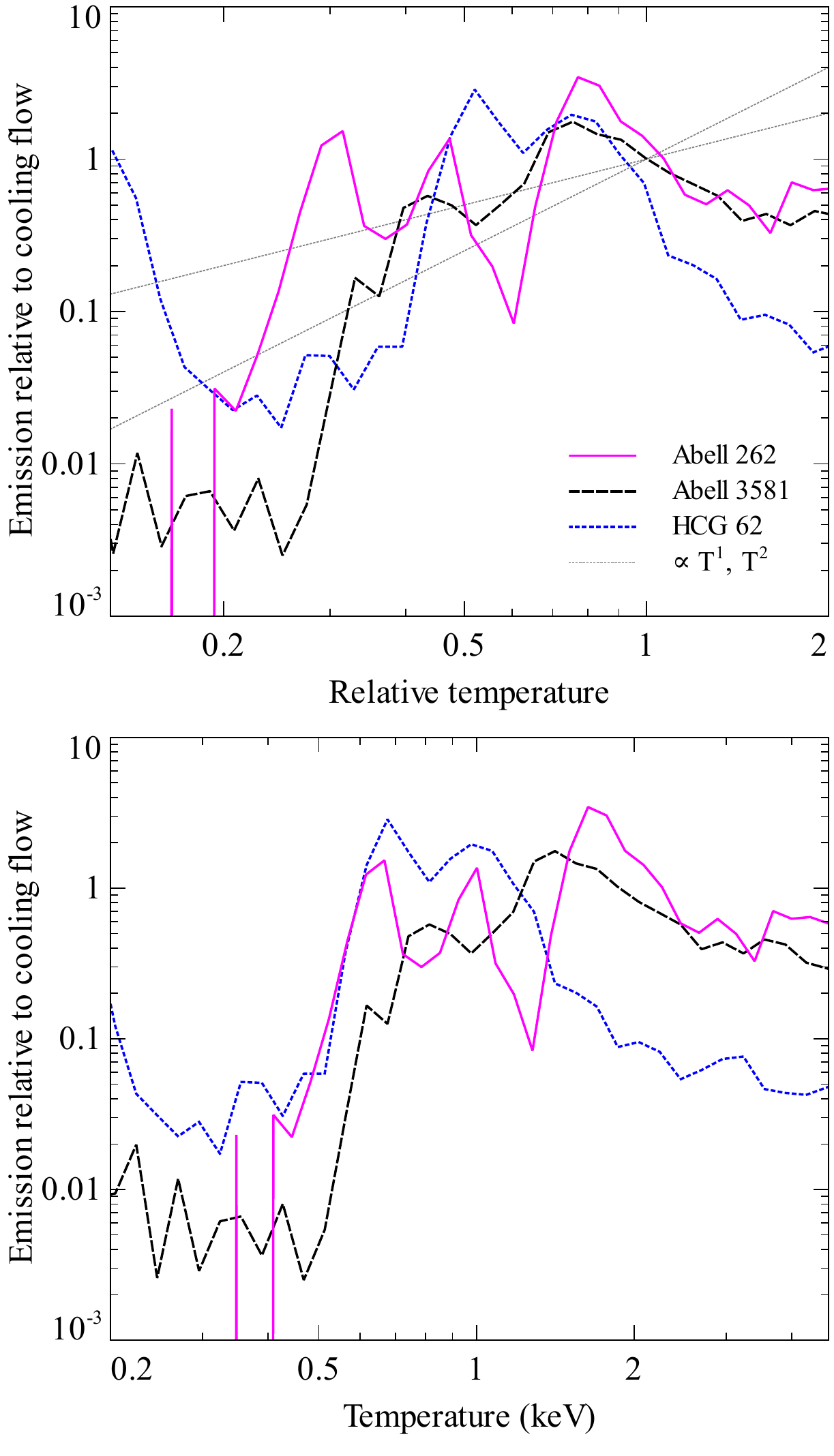}
 \caption{Distribution of gas relative to the amount expected from a cooling flow model. The cooling flow mass deposition rates used were taken from the surface brightness deprojection results at the cooling radius (listed in Table~\ref{tab:deproj}). The top panel shows the gas distribution as a function of temperature relative to the temperature at the cooling radius. Powerlaw distributions of indices 1 and 2 are shown for comparison. The bottom panel shows the distribution as a function of absolute temperature.}
 \label{fig:tdist_rel_cflow}
\end{figure}

\cite{Peterson03} found that the differential luminosity distribution of gas as a function of temperature in a sample of clusters showed a powerlaw index of 1--2 up to an upper temperature. To examine whether this holds in these objects we show the distribution of gas as a function of temperature relative to a cooling flow model in Fig.~\ref{fig:tdist_rel_cflow}. To calculate this, we took the gas normalization-temperature distribution from the smooth particle analysis (Fig.~\ref{fig:smoothparticle}) and divided the normalizations by those from a cooling flow model at each cluster redshift. The cooling flow model was calculated with Solar metallicity and with the mass deposition rate taken from the \emph{Chandra} surface brightness deprojection at the cooling radius (listed in Table~\ref{tab:deproj}). The top panel in Fig.~\ref{fig:tdist_rel_cflow} shows the distribution as a function of temperature relative to the temperature at the cooling radius. The bottom panel shows the distribution as a function of absolute temperature.

In these objects the distribution appears more complex than a simple powerlaw distribution with an index of between 1 and 2. If plotted at absolute temperature there appears to be a sharp break at around 0.5~keV, with much less gas below this temperature. This break is consistent with the results from a cooling flow model with a non-zero minimum temperature (Section \ref{sect:cflowmodels}). A powerlaw is consistent with the results above this temperature. The results of the cluster sample in \cite{Peterson03} have no detections of gas below 0.5~keV (in figure 6 in that paper). These results would also be consistent with a powerlaw distribution above 0.5~keV and a break below that temperature, although the data quality for many of the objects makes this hypothesis difficult to test here. A temperature break around 0.5~keV was also seen in M87 \citep{Werner06M87,Simionescu08}.

If the breaks in the distributions around 0.5~keV are widespread, this could indicate that an extra physical process becomes relevant below around 0.5~keV temperature. This could, for example, be the onset of non-radiative cooling (see Section \ref{sect:nonradiative} below). Such a break would be easier to see in cooler objects such as we are examining in this paper. At this threshold temperature the X-ray emitting gas has cooling times less than $10^8$~yr.

\subsection{Heating the coolest gas}
All three of these clusters show signs of AGN activity in their cores. If AGN activity is supplying the correct amount of heat to prevent cooling, the problem is how can the AGN provide the right amount of heat to the coolest X-ray gas, which only occupies a small fraction of the core of the cluster? This material has the shortest X-ray cooling time and lowest temperature. Any sort of heating mechanism which heats in a volume-averaged way should not work. The heat would need to be targeted to this X-ray coolest gas in order to prevent that cooling.

The cool blobs are around half the temperature of their surroundings so their emissivity is about four times larger (the cooling curve is approximately flat over this temperature range). They therefore require a heating rate four times larger than their surrounds. If there is widespread heating per unit volume then the surroundings would be overheated to prevent the coolest blobs from cooling further still.

The material we observe in these cool clusters and groups is only a factor of 2.6 to 5 times smaller than the gas further out. The heating tuning problem is even worse for other objects. As mentioned in the introduction, there is cool gas a factor of eight smaller than the outskirts in 2A~0335+096 \citep{Sanders2A033509}, fifteen in Abell~2204 \citep{SandersA220409}, ten in Centaurus \citep{SandersRGS08}, and twelve in the filaments in Perseus \citep{FabianPer06}.

\subsection{An alternative to heating: non-radiative cooling}
\label{sect:nonradiative}
If AGNs are responsible for the bulk of the heating in clusters, then either the cool gas must be heated by other means, a mechanism is in place for channeling AGN heating to this cool gas, or there are significant amounts cooling in the central regions. Significant amounts of cooling are possible if the gas is not cooling radiatively. For instance, mixing of X-ray emitting material with cooler gas would allow the material to cool much quicker \citep{FabianCFlow02}. Energetically, the amount of energy lost in the cooling could easily be lost in the infrared output of a central cluster galaxy.

The missing luminosity in X-rays for $10\Msunpyr$ cooling from 1~keV at $0.6\Zsun$ is around $2\times10^{42} \ergps$. Abell~262 has an infrared luminosity of $8 \times 10^{42} \ergps$ \citep{ODea08}, compatible with the energy which would need to be emitted if cooling were taking place by mixing. 

Infrared emission is also seen from HCG~62 \citep{Johnson07} with a peak $24\mu$m flux of around 10~mJy from the galaxy at the X-ray centroid, NGC 4778. If the 8 and 24~$\mu$m fluxes are linearly interpolated to get the flux at 15~$\mu$m, the infrared luminosity of the galaxy can be estimated to be $1.4\times10^{43}\ergps$, using equation 13 of \cite{Elbaz02}. This is again, much more than the energy that would need to be emitted if the X-ray gas was cooling by mixing.

In objects such as Perseus and Centaurus, cool X-ray emitting filaments are associated with material at still colder temperatures and H$\alpha$ nebulosity (e.g. \citealt{FabianPerFilament03}, \citealt{Crawford05}, \citealt{Salome06}). In 2A~0335+096 H$\alpha$ emission is also seen associated with the X-ray cool blobs \citep{Sanders2A033509}. The cool X-ray emission and colder material must be intimately related.

\subsection{Cooler gas and star formation}
The infrared luminosity in Abell~262 can be converted to a star formation rate of $0.5 \Msunpyr$ \citep{ODea08}. If these stars were forming constantly at steady state from cooling X-ray gas, it is consistent with our lowest mass deposition rate upper limits. The cluster also has an H$\alpha$ emission with luminosity of $1.1\times10^{40}\ergps$ \citep{Hatch07}. In addition $4\times10^{8}\Msun$ of H$_2$ gas was detected in the central cluster galaxy \citep{Prandoni07}. Both the ionised and molecular gas have similar velocity structures, appearing to be emitted from a rotating disc.

Abell~3581 has extended H$\alpha$ emission \citep{DanzigerFocardi88}. HCG~62 shows very weak H$\alpha$ emission \citep{Spavone06}. \cite{Valluri96} calculate that the H$\alpha$ emission relative to the continuum is smaller than for other traditional cooling flow objects.

\subsection{Mass deposition rates}
Like the previous analyses of RGS spectra of cool core clusters, we find significantly less gas at low temperatures than would be expected if cooling were taken place at a steady state in the absence of heating. The best example of this is HCG~62, where our limits in the 0.24 to 0.49 keV temperature range are two orders of magnitude lower than the cooling rates compatible with the surface brightness profile or spectra of the hot gas. In Abell 262 and Abell 3581 the spectra are consistent with higher fractional rates of cooling.

The picture is made more complex by the discrepancy between the different plasma models over the strength of the Fe~\textsc{xvii} emission lines (Fig.~\ref{fig:lineratio}) and generally in spectral fitting (Figures \ref{fig:6cmpt_em}, \ref{fig:vmcflow_apec_spex} and \ref{fig:vmcflow_6_9}). In addition the iron and oxygen metallicities have some effect (Fig.~\ref{fig:omdot}; figure 17 in \citealt{SandersRGS08}). If, for some reason, the coolest gas is metal poor, the cooling rate may be being underestimated. Unless they are very underabundant in metals, this probably could not account for the very low cooling rate seen in HCG~62.

\section{Conclusions}
The deep \emph{XMM-Newton} RGS spectra of these objects show Fe~\textsc{xvii} line emission, indicating material around 0.5~keV in temperature. The width of the emission lines and the spatial extent of the cold gas seen by \emph{Chandra} show that the regions they are emitted from are resolved. Assuming pressure equilibrium, we can see that this cool component is not volume filling in Abell~262 and Abell~3581 and likely consists of cool blobs such as those seen in 2A~0335+096 or filaments as in Perseus or Centaurus.

The strength of these Fe~\textsc{xvii} emission lines and lack of O~\textsc{vii} emission lines show that there is substantially less material seen at lower temperatures than expected in the cooling flow picture, as seen in previous studies.

The multiphase nature of this cold material presents problems for heating mechanisms which are preventing the bulk of the material from cooling. The coolest material requires four times more heat than its immediate surroundings. Any heating mechanism stopping this material from cooling needs to target it in some way, or it is cooling, perhaps non-radiatively. Non-radiative cooling, perhaps by mixing, could be important at the lowest temperatures. In Abell~262 and HCG~62 the infrared fluxes of the central galaxy could contain the heat output of this process.

\section*{Acknowledgements}
ACF thanks the Royal Society for support.  JRP and KF are supported by NASA grant \#NNX07AQ30G. HRR acknowledges the support of the Science and Technology Facilities Council. We thank the anonymous referee for helpful comments.

\bibliographystyle{mnras}
\bibliography{refs}

\clearpage
\end{document}